\documentclass[paper]{JHEP3}
\pdfoutput=1
\usepackage{amsmath,amssymb,amsthm,amscd,graphicx}
\input epsf.sty

\theoremstyle{definition}

\newcommand{\CL}{{\cal L}}
\newcommand{\CM}{{\cal M}}
\newcommand{\CN}{{\cal N}}
\newcommand{\CO}{{\cal O}}

\newcommand{\CT}{{\cal T}}

\newcommand{\CW}{{\cal W}}

\def\IZ{{\mathbb Z}}
\def\IR{{\mathbb R}}
\def\IC{{\mathbb C}}
\def\IP{{\mathbb P}}
\def\IT{{\mathbb T}}
\def\IS{{\mathbb S}}

\def\IN{{\mathbb N}}

\newcommand{\re}{{\rm e}}
\newcommand{\ri}{{\rm i}}
\newcommand{\rd}{{\rm d}}

\def\l{\ell}
\newcommand{\nn}{\nonumber \\}
\def\({\Bigl(}
\def\){\Bigr)}

\newcommand{\be}{\begin{equation}}
\newcommand{\ee}{\end{equation}}
\newcommand{\ba}{\begin{aligned}}
\newcommand{\ea}{\end{aligned}}
\newcommand{\ben}{\begin{eqnarray}\displaystyle}
\newcommand{\een}{\end{eqnarray}}

\newcommand{\sectiono}[1]{\section{#1}\setcounter{equation}{0}}

\newdimen\tableauside\tableauside=1.0ex
\newdimen\tableaurule\tableaurule=0.4pt
\newdimen\tableaustep
\def\phantomhrule#1{\hbox{\vbox to0pt{\hrule height\tableaurule width#1\vss}}}
\def\phantomvrule#1{\vbox{\hbox to0pt{\vrule width\tableaurule height#1\hss}}}
\def\sqr{\vbox{%
  \phantomhrule\tableaustep
  \hbox{\phantomvrule\tableaustep\kern\tableaustep\phantomvrule\tableaustep}%
  \hbox{\vbox{\phantomhrule\tableauside}\kern-\tableaurule}}}
\def\squares#1{\hbox{\count0=#1\noindent\loop\sqr
  \advance\count0 by-1 \ifnum\count0>0\repeat}}
\def\tableau#1{\vcenter{\offinterlineskip
  \tableaustep=\tableauside\advance\tableaustep by-\tableaurule
  \kern\normallineskip\hbox
    {\kern\normallineskip\vbox
      {\gettableau#1 0 }%
     \kern\normallineskip\kern\tableaurule}%
  \kern\normallineskip\kern\tableaurule}}
\def\gettableau#1{\ifnum#1=0\let\next=\null\else
\squares{#1}\let\next=\gettableau\fi\next}

\newcommand{\figref}[1]{Fig.~\protect\ref{#1}}

\preprint{DESY\ 13-096, TIT/HEP-627}

\title{Non-perturbative effects and the refined topological string}

\author{
Yasuyuki Hatsuda$^a$, Marcos Mari\~no$^b$, Sanefumi Moriyama$^c$ and Kazumi Okuyama$^d$
\\
$^a$DESY Theory Group, DESY Hamburg,\\ 
Notkestrasse 85, D-22603 Hamburg, Germany\\
and\\
Department of Physics,\\
Tokyo Institute of Technology, Tokyo 152-8551, Japan\\
\\
$^b$D\'epartement de Physique Th\'eorique et Section de Math\'ematiques,\\
Universit\'e de Gen\`eve, Gen\`eve, CH-1211 Switzerland\\
\\
$^c$Kobayashi Maskawa Institute and Graduate School of Mathematics,\\
Nagoya University, Nagoya 464-8602, Japan\\
\\
$^d$Department of Physics,\\
Shinshu University, Matsumoto 390-8621, Japan\\

\email{yasuyuki.hatsuda@desy.de, marcos.marino@unige.ch,
moriyama@math.nagoya-u.ac.jp, kazumi@azusa.shinshu-u.ac.jp}}

\abstract{The partition function of ABJM theory on the three-sphere has non-perturbative corrections due to membrane instantons in the M-theory dual. We show that 
the full series of membrane instanton corrections is completely determined by the refined topological string on the Calabi--Yau manifold 
known as local $\IP^1 \times \IP^1$, in the Nekrasov--Shatashvili limit. Our result can be interpreted as a first-principles 
derivation of the full series of non-perturbative effects for the closed topological string on this Calabi--Yau background. 
Based on this, we make a proposal for the non-perturbative free energy of topological 
strings on general, local Calabi--Yau manifolds. }

\begin{document}

\sectiono{Introduction}

Large $N$ dualities relate gauge theories to string theories, and provide in principle a non-perturbative definition 
of string theory on certain backgrounds. The genus expansion of string theory amplitudes emerges then as an asymptotic, $1/N$ expansion of gauge theory amplitudes. 
Most of the work on large $N$ dualities has focused on the large $N$ or planar limit of the correspondence. One can also use these dualities to 
extract information about subleading $1/N$ corrections, although this is typically more difficult and it has been comparatively much less explored. 
In principle, large $N$ dualities could be also used to study non-perturbative stringy effects, which correspond to corrections which are exponentially suppressed as $N$ becomes large. 
Results along this direction have been even rarer.

In this paper we use large $N$ dualities to completely determine the non-perturbative structure of the free energy of M-theory on 
AdS$_4 \times \IS^7/\IZ_k$. As a bonus, we obtain as well the non-perturbative structure for the free energy of topological string theory on the Calabi--Yau manifold 
known as local $\IP^1 \times \IP^1$, since both problems are formally identical. The non-perturbative structure we find 
turns out to be encoded by the {\it refined} topological string on local $\IP^1 \times \IP^1$, in the so-called Nekrasov--Shatashvili (NS) limit \cite{ns}. 

The solution to this problem has been based on the convergence of many different results. First of all, a large $N$ dual to M-theory on AdS$_4 \times \IS^7/\IZ_k$ 
was proposed already in \cite{maldacena} in terms of the theory of $N$ coincident M2 branes. In \cite{abjm}, based on previous work \cite{review}, 
this theory was constructed as an $\CN=6$ 
supersymmetric $U(N) \times U(N)$ Chern--Simons--matter theory known as ABJM theory. 
In this large $N$ duality, the geometric parameter $k$ in M-theory 
corresponds to the Chern--Simons coupling. The second ingredient was the localization computation of \cite{kwy}, where the partition function of ABJM theory 
on the three-sphere was reduced to a matrix integral which we will call the ABJM matrix model. This matrix model has been intensively studied from many points of view, and a variety of results have been found. The planar free energy, as well as the subleading $1/N$ corrections in the standard 't Hooft or genus expansion, were determined in \cite{dmp}. This expansion makes contact with the type IIA reduction of M-theory and it captures all worldsheet instanton corrections to the partition function. However, in order to make contact with the M-theory regime, one should study the ABJM matrix model in the so-called M-theory expansion, 
where $N$ is large but $k$ is fixed. This was first done in \cite{hkpt}, where the leading, large $N$ limit was studied. In order to understand in more detail the M-theory expansion, and the corrections to the large $N$ limit, a new method was introduced in \cite{mp-fermi}, based on an equivalence with an ideal Fermi gas. In this approach, the Planck constant 
of the quantum gas is naturally identified with the inverse string coupling, and the semiclassical limit of the gas corresponds then to the strong string coupling limit in 
type IIA theory. One of the main virtues of the Fermi gas approach is that it makes it possible to calculate systematically non-perturbative stringy effects. These effects were anticipated in \cite{dmpnp}, where they were interpreted as membrane instanton effects in M-theory, or equivalently as D2-brane effects in type IIA theory. Thus, the Fermi gas approach opened the way for a quantitative determination of these effects in the 
M-theory dual to ABJM theory.

During the last year, the Fermi gas approach has led to many results on the partition function of ABJM theory. The equivalence between this method and the 
TBA system of \cite{zamo,tw} has been particularly useful. We now have a lot of data, like for example WKB expansions at small $k$ 
of the membrane instanton corrections \cite{mp-fermi,cm}. The calculation of the values of the partition function for various values of $N$ and $k$ \cite{hmo,py,hmo1}, 
and their extrapolation to large $N$, have produced numerical results for the exponentially small corrections. 
In \cite{hmo1,hmo2}, it was noticed that the corrections due to worldsheet instantons, which are known explicitly, are singular for integer values of $k$. 
Since the partition function is regular for all $k$, it was postulated that these singularities should be cancelled by membrane instanton corrections, 
as well as corrections coming from bound states of 
membranes and fundamental strings. This principle, which we will call the HMO cancellation mechanism, when combined with 
WKB expansions and numerical results, has led to conjectural exact results in $k$ for the very first membrane instanton 
corrections \cite{hmo1,cm,hmo2} and to a conjecture for the structure of bound states \cite{hmo2}. According to this conjecture, the bound states are completely determined by 
the worldsheet instantons and the membrane instanton corrections. The remaining open problem is then to find an analytic description of the membrane instanton corrections in the M-theory regime, 
i.e. as an expansion at large $N$ but exact in $k$. 

In this paper we find precisely such a description. It turns out that the membrane instanton expansion at large $N$, which involves two independent generating functionals, is completely determined 
by the NS limit of the refined topological string on local $\IP^1\times \IP^1$. This limit is described by the two quantum periods of the mirror manifold \cite{mm1,mm2,acdkv}, 
which are equal to the two generating functionals we were looking for. The Chern--Simons coupling $k$ of ABJM theory corresponds to the quantum deformation parameter $\hbar$, and the standard large radius expansion of the periods corresponds precisely to the large $N$ expansion in ABJM theory. Since the periods can be calculated exactly as a function of $\hbar$, this equivalence solves the problem of computing the non-perturbative corrections to the free energy of ABJM theory.  

So far we are lacking a proof of this equivalence, which we have checked by comparing the existing results on membrane instantons in ABJM theory to the explicit results for the quantum periods, so our result here should be regarded as a conjecture. It can be stated quite precisely as an equivalence between the solution of the TBA system describing the ABJM partition function which is analytic at $k=0$, and the problem of quantizing the 
periods of local $\IP^1 \times \IP^1$. 

One of the first insights which made possible a precise quantitative understanding of the ABJM matrix model is its equivalence \cite{mp} to the matrix model describing 
Chern--Simons theory on $\IR\IP^3$ \cite{mm}, which is dual at large $N$ to topological string theory on local $\IP^1 \times \IP^1$ \cite{akmv-cs}. This implies, for example, that 
the worldsheet instanton corrections in ABJM theory are determined by the worldsheet instanton corrections in this topological string theory. We can then {\it define} the non-perturbative 
partition function of topological string on local $\IP^1 \times \IP^1$ through the ABJM matrix model. With this non-perturbative definition, our computation of exponentially small corrections to this matrix model partition function 
can be also regarded as a derivation of the full structure of non-perturbative effects for topological string theory on 
local $\IP^1 \times \IP^1$. The fact that the Fermi gas approach could be used to obtain a precise quantitative understanding of 
non-perturbative effects in this topological string model was pointed out in \cite{mp-fermi}, and emphasized in \cite{mytalk}. 

The non-perturbative structure of topological strings has been the subject of much speculation in recent years, and there are by now various proposals on how it should look like. We would like to emphasize, however, that our derivation of the non-perturbative structure in this particular example is done from first principles, once we define it through the large $N$ matrix model dual, 
and it fits a large amount of data on the large $N$ asymptotics of the matrix model. 
Our result says that the non-perturbative part of the standard topological string free energy is determined by the refined topological string in the NS limit, 
on the same background. Inspired by this concrete result, we make a proposal for the non-perturbative structure of topological 
strings on arbitrary local CY manifolds, where the non-perturbative effects are encoded in the refined topological string. 
It turns out that our proposal (as well as our concrete, first-principles calculation for local $\IP^1 \times \IP^1$) is similar to a recent proposal by 
Lockhart and Vafa \cite{lv}, which was inspired by localization in five-dimensional supersymmetric Yang--Mills theories, and we point out the resemblances as well as the differences between the 
two proposals. 

The organization of this paper is as follows. In section 2 we review the known results on the 
grand potential of ABJM theory obtained in \cite{mp-fermi, hmo,py,hmo1,cm,hmo2}. In section 3 we show that these results 
are encoded in the NS limit of the refined topological string, and in particular in the quantum periods. In section 4 we point out that this leads 
to the determination of the non-perturbative structure of the topological string on local $\IP^1 \times \IP^1$, and we make a 
proposal on how to extend this to arbitrary, local CY manifolds. We also discuss the relationship of our results and proposal to 
the work of \cite{lv}. Finally, in section 5 we conclude and discuss some avenues for further research. 
In Appendix A we explain how to calculate the quantum A-periods from the TBA system of the Fermi gas, and in Appendix B we make some comments on the quantum mirror map.

\sectiono{The partition function of ABJM theory}

\subsection{The grand potential}
As it was shown in \cite{kwy}, the partition function of ABJM theory on the three-sphere, $Z(N,k)$, is given by the matrix integral
\be
\label{abjmmatrix}
\ba
&Z(N,k)\\
&={1\over N!^2} \int {\rd ^N \mu \over (2\pi)^N} {\rd ^N \nu \over (2\pi)^N} {\prod_{i<j} \left[ 2 \sinh \left( {\mu_i -\mu_j \over 2} \right)\right]^2
  \left[ 2 \sinh \left( {\nu_i -\nu_j \over 2} \right)\right]^2 \over \prod_{i,j} \left[ 2 \cosh \left( {\mu_i -\nu_j \over 2} \right)\right]^2 } 
  \exp \left[ {\ri k \over 4 \pi} \sum_{i=1}^N (\mu_i^2 -\nu_i^2) \right].
  \ea
  \ee
This matrix integral can be calculated in two different regimes. In the {\it 't Hooft expansion} one considers the limit
\be
\label{thooftl}
N \rightarrow \infty, \quad \lambda={N\over k}\, \,\,  \text{fixed},
\ee
and the partition function has the standard $1/N$ expansion, 
\be
\label{genux-ex}
Z(N,k)=\exp \left[ \sum_{g=0}^{\infty} N^{2-2g} F_g(\lambda)\right],
\ee
which corresponds to the genus expansion of type IIA superstring theory on AdS$_4\times \IC\IP^3$ \cite{abjm}. 
The genus $g$ free energies $F_g(\lambda)$ can be calculated exactly as a function of $\lambda$, and order by order in the genus expansion, by using matrix model 
techniques \cite{dmp}. They contain 
non-perturbative information in $\alpha'$, since they involve exponentially small corrections of the form 
\be
\CO\left(\re^{-2 \pi {\sqrt{2\lambda}}}\right).
\ee
It was conjectured in \cite{dmp} that these terms correspond to worldsheet instantons wrapping a two-cycle $\IC\IP^1 \subset \IC\IP^3$, which were first considered in \cite{cagnazzo}. 

In the {\it M-theory expansion}, one computes the partition function in the regime
\be
\label{mthl}
N \rightarrow \infty, \quad k\, \,\,  \text{fixed}.
\ee
This is the regime which is suitable for the dual description in terms of M-theory on AdS$_4\times \IS^7/\IZ_k$. In this regime, one expects to 
find as well non-perturbative effects in the string coupling constant, 
which in type IIA theory correspond to Euclidean D2-brane instantons wrapping three-cycles in the target space. 
In \cite{dmpnp} an appropriate, explicit family of generalized Lagrangian submanifolds with the topology of $\IR\IP^3\subset \IC\IP^3$ was proposed as an explicit candidate for this type of cycles, leading to exponentially small corrections of the form 
\be
\label{membranea}
\exp\left( -k \pi {\sqrt{2 \lambda}} \right) .
\ee

In order to understand the M-theory expansion of the ABJM matrix integral, one needs a suitable approach, different from the standard $1/N$ expansion of matrix integrals. 
A first step in this direction 
was taken in \cite{hkpt}, where the leading contribution to the partition function at large $N$ and fixed $k$ was determined for various ${\cal N}=3$ Chern--Simons--matter theories. 
A more systematic approach to the problem was introduced in
\cite{mp-fermi}, and it is based on an analogy to a quantum, ideal Fermi gas. One first notices (see also \cite{kwytwo}) that the matrix integral (\ref{abjmmatrix}) can be written as 
\be
\label{fg-matrix}
Z(N,k)={1 \over N!} \sum_{\sigma  \in S_N} (-1)^{\epsilon(\sigma)}  \int  {\rd ^N x \over (2 \pi k)^N} {1\over  \prod_{i} 2 \cosh\left(  {x_i  \over 2}  \right)
2 \cosh\left( {x_i - x_{\sigma(i)} \over 2 k} \right)}.
\ee
This in turn can be interpreted as the canonical partition function of a one-dimensional Fermi gas with a non-trivial one-particle density matrix
\be
\label{densitymat}
\rho(x_1, x_2)={1\over 2 \pi k} {1\over \left( 2 \cosh  {x_1 \over 2}  \right)^{1/2} }  {1\over \left( 2 \cosh {x_2  \over 2} \right)^{1/2} } {1\over 
2 \cosh\left( {x_1 - x_2\over 2 k} \right)}.
\ee
The one-particle Hamiltonian $\hat H$ of this system is then defined as
\be
\label{onepH}
\hat \rho=\re^{-\hat H}, \qquad \langle x_1 | \hat \rho | x_2 \rangle = \rho(x_1, x_2), 
\ee
and the Planck constant of the Fermi gas is 
\be
\label{planck-fermi}
\hbar_{\rm FG} = 2\pi k. 
\ee
The semiclassical or WKB expansion is then around $k=0$, and it corresponds to the strong string coupling expansion in the type IIA dual. The Fermi gas approach 
makes it possible to determine both the subleading $1/N$ corrections and non-perturbative corrections due to D2-brane instantons. Various aspects of this approach have been 
developed in \cite{hmo,py,kmss, hmo,hmo2,ahs} and we will review some of them in this section. 

The Fermi gas approach suggests to look instead to the grand partition function (see also \cite{okuyama})
\be
\label{grand}
\Xi(\mu, k)=1+\sum_{N=1}^\infty Z(N,k) z^N, 
\ee
where 
\be
z=\re^{\mu}
\ee
plays the r\^ole of the fugacity and $\mu$ is the chemical potential. The grand potential is then defined as
\be
J(\mu,k) =\log \Xi (\mu, k). 
\ee
The canonical partition function is recovered from the grand-canonical potential as
\be
\label{exactinverse}
Z(N, k) =\oint {\rd z \over 2 \pi \ri } {\Xi (\mu, k) \over z^{N+1}}.
\ee
As explained in \cite{hmo1}, the grand potential has a ``naive" part, which is the one obtained with the standard 
techniques in Statistical Mechanics, and an oscillatory part which restores the $2 \pi \ri$ periodicity in $\mu$. It turns out that the contour 
in (\ref{exactinverse}) can be deformed to the imaginary axis if one replaces the grand potential by its ``naive" part, which will be the only one we will consider in this 
paper. Therefore, we can write
\be
\label{muint}
Z(N,k) ={1\over 2 \pi \ri} \int_{-\ri \infty}^{\ri \infty} \rd \mu \, \exp\left[J(\mu,k) - \mu N\right], 
\ee
and compute $J(\mu, k)$ with standard techniques. 

As shown in \cite{mp-fermi}, the grand potential is the sum of a perturbative and a non-perturbative piece, 
\be
J(\mu, k)=J^{(\rm p)}(\mu, k) + J^{(\rm np)}(\mu, k). 
\ee
The perturbative piece is a cubic polynomial in $\mu$: 
\be
J^{(\rm p)}(\mu,k)= {C(k) \over 3} \mu^3 + B(k) \mu + A(k), 
\ee
where 
\be
\label{CB}
C(k)= {2\over \pi^2 k}, \qquad B(k)={k \over 24} + {1\over 3k}.
\ee
The coefficient $A(k)$ can be computed in a WKB expansion around $k=0$ \cite{mp-fermi}, and the all-orders result was 
conjectured in \cite{hanada}. When inserted in (\ref{muint}), the perturbative piece $J^{(\rm p)}(\mu,k)$ leads to the Airy function result for $Z(N,k)$ first obtained in \cite{fhm}.

\subsection{The structure of the non-perturbative corrections}

\FIGURE{
\includegraphics[height=6cm
]{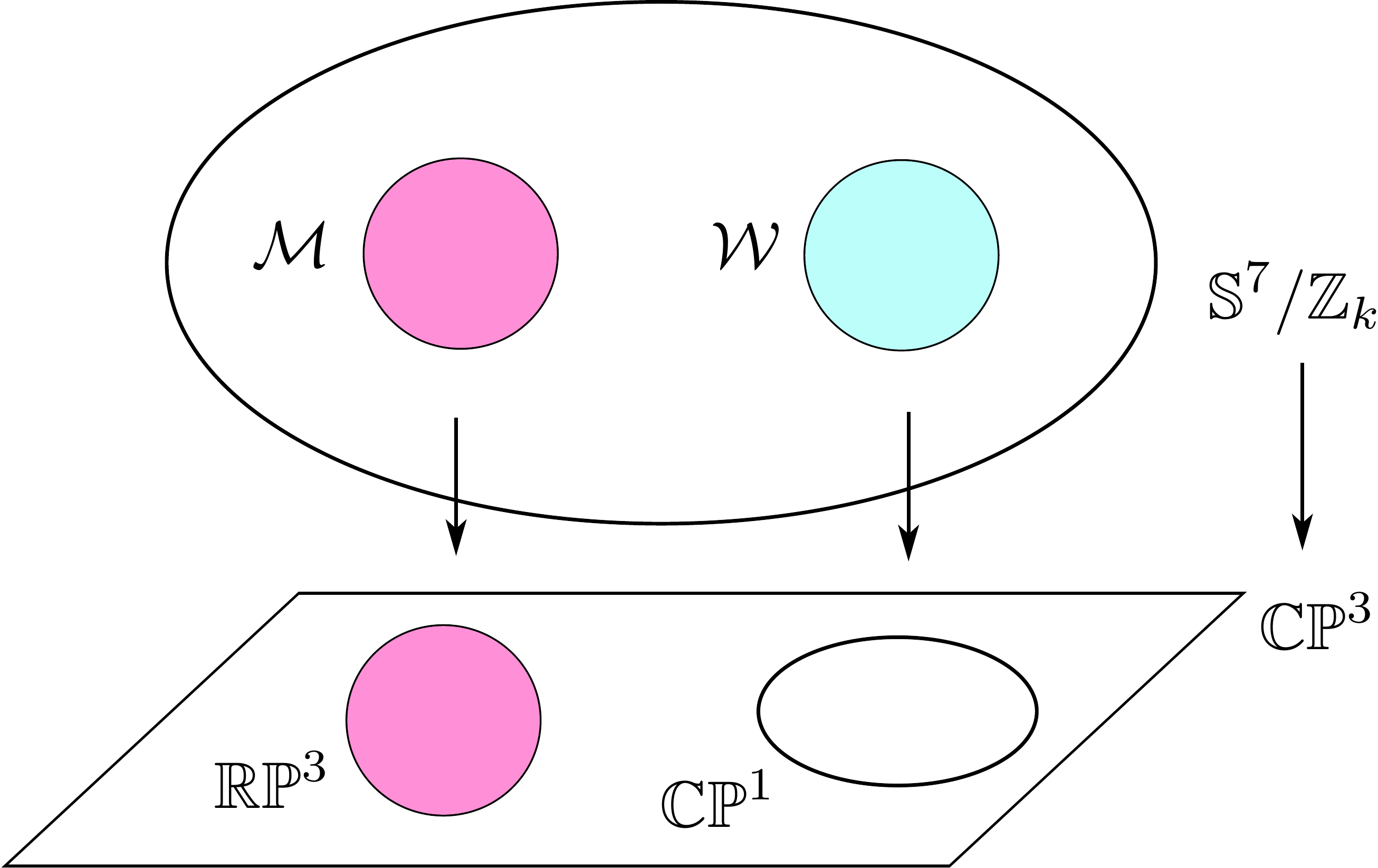} 
\caption{From the M-theory point of view, there are two types of non-perturbative effects in ABJM theory: 
M2-branes can wrap a cycle $\CM \subset \IS^7/\IZ_k$ which descends to an $\IR\IP^3 \subset \IC\IP^3$ cycle 
in the type IIA target; or they can wrap a cycle $\CW$ which descends to $\IC\IP^1\subset \IC\IP^3$. 
The most general M2-brane configuration wraps $\ell$ times the cycle $\CM$ and $m$ times the cycle $\CW$.}
\label{in-fig}
}

In this paper we will be interested in the 
non-perturbative part of the grand potential $J^{\rm np}(\mu,k)$. This function encodes the exponentially small, 
non-perturbative corrections at large $N$ to the matrix model (\ref{abjmmatrix}). It has the following expansion 
at fixed $k$ and large $\mu$:
\be
\label{jnpgen}
J^{\rm (np)}(k,\mu)
=\sum_{\substack{\l,m=0 \\ (\l,m) \ne (0,0)}}^\infty
f_{\l, m}(k,\mu) \exp\left[-\left( 2\l+\frac{4m}{k}\right)\mu\right], 
\ee
where $\ell, m$ are non-negative integers. This structure is the expected one from the point of view of M-theory. Indeed, as it is well-known \cite{bbs}, 
in M-theory both worldsheet instantons and D2-brane instantons get unified in terms of M2-branes wrapping three-cycles. In our case, D2-brane instantons correspond to 
a three-cycle $\CM \subset \IS^7/\IZ_k$ which descends to a three-cycle in type IIA theory; worldsheet instantons correspond to M2-branes wrapping the three-cycle $\CW=\IS^3/\IZ_k$, which descends to the two-cycle $\IC\IP^1\subset \IC\IP^3$. Therefore, from the point of view of M-theory, there are two types of three-cycles, $\CM$ and $\CW$, and the most general M2-brane configuration wraps $\ell$ 
times the cycle $\CM$ and $m$ times the cycle $\CW$, see \figref{in-fig}. The term labelled by the integers $(\ell,m)$ in (\ref{jnpgen}) gives the contribution of such a bound state. 
The large $\mu$ expansion (\ref{jnpgen}), after plugged in (\ref{muint}), leads to the asymptotic expansion of $Z(N,k)$ at large $N$. Since, at large $N$, the Legendre transform of $J(\mu)$ sets the value of $\mu$ to be at the saddle-point 
\be
\mu \approx  {\sqrt{N \over C(k)}}, 
\ee
the exponential in (\ref{jnpgen}) gives the correct weight for a bound state of $\ell$ worldsheet instantons and $m$ D2-brane instantons in $Z(N,k)$. 

To determine the full non-perturbative grand potential, and therefore the full non-perturbative structure of the ABJM partition function, we have to find a systematic way of computing the coefficients $f_{\l, m}(k,\mu)$. Let us now review what is known about them. 

\subsubsection{Worldsheet instantons and the topological string}

The contribution to (\ref{jnpgen}) with $\ell=0$ will be denoted by $J^{\rm WS}(\mu, k)$, and it is due to worldsheet 
instantons wrapping $\IC\IP^1 \subset \IC\IP^3$.
It contains the same information than the genus $g$ free energies
$F_g(\lambda)$ appearing in the 't Hooft expansion, but reorganized in a different way, since in the M-theory expansion $k$ is fixed.
It is possible to write down an explicit expression for $J^{\rm WS}(\mu, k)$ by relating it to topological string theory. Let us first recall some results from topological string theory. 
If $X$ is a CY manifold with K\"ahler parameters $T_I$, $I=1, \cdots, n$, the free energy of topological string theory in the large radius frame has the following form \cite{gv}:
\be
\label{gv-exp}
F({\bf Q}, g_s)= \sum_{g\ge0}\sum_{w\ge 1} \sum_{{\bf d}} 
{(-1)^{g-1} \over w} n _g^{{\bf d}} \left( q_s^{w/2} -q_s^{-w/2} \right)^{2g-2} {\bf Q}^{w {\bf d}}.
\ee
In this formula, 
\be
q_s=\re^{g_s},
\ee
where $g_s$ is the topological string coupling constant, and we have denoted
\be
{\bf d}=(d_1,\cdots, d_n), \qquad Q_I =\re^{-T_I}, \quad I=1, \cdots, n, 
\ee
as well as 
\be
{\bf Q}^{\bf d}= Q_1^{d_1} \cdots Q_n^{d_n}.
\ee
The integer numbers $n _g^{{\bf d}}$ are the Gopakumar--Vafa (GV) invariants of $X$ at genus $g$ and degrees ${\bf d}=(d_I)$. 
 
 The relevant manifold for the partition function of ABJM theory is the non-compact CY known as local 
 $\IP^1 \times \IP^1$, which is the total space of the anti-canonical bundle over the surface $\IP^1 \times \IP^1$. This 
 space has two K\"ahler moduli $T_1$, $T_2$, corresponding to the two $\IP^1$s, and GV invariants $n^{d_1, d_2}_g$. 
 In terms of these invariants, the contribution of worldsheet instantons to the grand potential $J^{\rm WS}(\mu, k)$ is given by \cite{hmo1}:
\be
\label{gvone}
J^{\rm WS}(\mu, k)= \sum_{g\ge 0} \sum_{w,d \ge1} n^d_g \left( 2 \sin {2 \pi w \over k} \right)^{2g-2} {(-1)^{d w}  \over w}{\rm e}^{-{4 d w \mu \over k}}.
\ee
In this formula, $n_d^g$ is the ``diagonal" GV invariant,
\be
\label{gvtwo}
n^d_g= \sum_{d_1+d_2 =d} n^{d_1, d_2}_g. 
\ee
This formula follows from the fact  the ABJM matrix integral computes the partition function of 
topological string theory of the ``diagonal" local $\IP^1 \times \IP^1$, where the two K\"ahler parameters are identified,
\be
\label{T-slice}
T_1=T_2=T,
\ee
and in the orbifold frame \cite{mp,dmp}. 
The grand potential $J(\mu)$ is related to this partition function by an inverse Legendre transform, 
and by using the results of \cite{abk} it is easy to see \cite{mp-fermi} that it 
is essentially given by the topological string free energy 
in the large radius frame. This topological string free energy $F(T, g_s)$ depends on the diagonal K\"ahler modulus $T$ in (\ref{T-slice}) and on the 
topological string coupling constant $g_s$, and by comparing (\ref{gv-exp}) with (\ref{gvone}) we have the relationship \cite{mp-fermi,hmo1}
\be
\label{relatio}
g_s= {4 \pi \ri \over k}, \qquad T= {4 \mu\over k} - \pi \ri. 
\ee
The formula (\ref{gvone}) reduces then the computation of the coefficient $f_{0, m}(k,\mu)$ in (\ref{jnpgen}) to the solvable problem of 
computing the GV invariants of local $\IP^1 \times \IP^1$. This can indeed be done by the topological 
vertex \cite{amv,akmv} or by direct integration of the holomorphic anomaly equations \cite{hkr,dmp}, for example. 

Notice that the Gopakumar--Vafa representation of the free energy is precisely what is needed for the M-theory expansion: it resums the genus expansion 
order by order in the exponentiated parameter $\re^{-T}$, therefore it leads to an expansion at large $N$ in ABJM theory, but which is {\it exact} in $k$ at each order in $\re^{-\mu/k}$. 

\subsubsection{Membrane instantons and bound states} 
The contribution to (\ref{jnpgen}) with $m=0$ will be denoted by $J^{\rm M2}(\mu, k)$, and it is due to M2-branes wrapping the three-cycle $\CM$. They lead to non-perturbative effects in the string coupling constant of the type IIA superstring. As shown by explicit calculation in \cite{mp-fermi}, $J^{\rm M2}(\mu, k)$ has the following expansion for $\mu \gg 1$, 
\be
\label{gen-M2}
J^{\rm M2}(\mu, k) = \sum_{\ell\ge 1} \left(a_\ell (k) \mu^2 + b_\ell (k)  \mu + c_\ell(k) \right) {\rm e}^{-2 \ell \mu}. 
\ee
This is of course an expansion at large $N$. We will refer to the $\ell$-th term in the infinite series (\ref{gen-M2}) as the contribution of the $\ell$-membrane instanton. The coefficients $a_\ell(k)$, $b_\ell(k)$ and $c_\ell(k)$ are non-trivial functions of $k$, and we will discuss how to compute them in the next subsection. 

The above results do not determine the contributions coming from bound states. A conjecture for their structure was put forward in \cite{hmo2}. 
Let us define an ``effective" chemical potential $\mu_{\rm eff}$:
\be
\label{mueff-mu}
\mu_{\rm eff}= \mu + {1\over C(k)} \sum_{\ell=1}^\infty a_\ell(k) \re^{-2\ell \mu}, 
\ee
where $C(k)$ is given in (\ref{CB}) and the coefficients $a_\ell(k)$ are the ones appearing in (\ref{gen-M2}).
Then, it was conjectured in \cite{hmo2} that the contribution of bound states is obtained by simply putting $\mu_{\rm eff}$ instead of $\mu$ in $J^{\rm WS}(\mu, k)$. 
In other words, according to this conjecture, $J^{\rm (np)}(\mu,k)$ is simply given by 
\be
\label{conj-one}
J^{\rm (np)}(\mu,k)= J^{\rm WS}(\mu_{\rm eff}, k)+J^{\rm M2}(\mu, k). 
\ee
Since the worldsheet instanton contribution is known, this conjecture reduces the calculation of the 
non-perturbative grand potential to the determination of the coefficients of the membrane instanton expansion (\ref{gen-M2}). 

Actually, another conjecture in \cite{hmo2} relates the coefficient $c_\ell(k)$ to the other two coefficients $a_\ell(k)$, $b_\ell(k)$. In order to write down this relationship in an elegant way, we introduce generating functionals for the three sets of coefficients in $J^{\rm M2}(\mu, k)$,   
\begin{align}
\label{Jabc}
J_a(\mu,k)=\sum_{\l=1}^\infty a_\l(k)\re^{-2\l \mu},\quad
J_b(\mu, k)=\sum_{\l=1}^\infty b_\l(k)\re^{-2\l \mu},\quad
J_c(\mu, k)=\sum_{\l=1}^\infty c_\l(k)\re^{-2\l \mu}.
\end{align}
We now write the perturbative and membrane instanton part of the grand potential in terms of the 
effective chemical potential $\mu_{\rm eff}$, as follows:
\begin{align}
J^{(\rm p)}(\mu,k)+J^{\rm M2}(\mu,k)&=J^{(\rm p)}(\mu_{\rm eff},k)+\mu_{\rm eff}
\widetilde{J}_b(\mu_{\rm eff},k)+\widetilde{J}_c(\mu_{\rm eff},k),
\end{align}
where $\widetilde{J}_b(\mu_{\rm eff},k)$ and
$\widetilde{J}_c(\mu_{\rm eff},k)$ are given by
\begin{align}
\label{widetildeJ}
\widetilde{J}_b(\mu_{\rm eff},k)
&
=J_b(\mu,k)-\frac{J_a(\mu,k)^2}{C(k)},\\
\widetilde{J}_c(\mu_{\rm eff},k)
&
=J_c(k,\mu)-\frac{J_a(\mu, k) J_b(\mu, k)}{C(k)}
-\frac{B(k)}{C(k)}J_a(\mu, k)+\frac{2J_a(\mu,k)^3}{3C(k)^2}.\notag
\end{align}
We conclude that the total grand potential of ABJM theory can be written as 
\be
\label{gpmueff}
J(\mu, k)=J^{(\rm p)}(\mu_{\rm eff},k)+J^{\rm WS} (\mu_{\rm eff},k)+ \mu_{\rm eff}
\widetilde{J}_b(\mu_{\rm eff},k)+\widetilde{J}_c(\mu_{\rm eff},k).
\ee
The two functions $\widetilde{J}_b(\mu_{\rm eff},k)$ and
$\widetilde{J}_c(\mu_{\rm eff},k)$, when expanded at large $\mu_{\rm eff}$, define the coefficients $\widetilde{b}_\l(k)$, $\widetilde{c}_\l(k)$:
\begin{align}
 \widetilde{J}_b(\mu_{\rm eff},k)=\sum_{\l=1}^\infty\widetilde{b}_\l(k)\re^{-2\l\mu_{\rm eff}}, \qquad \widetilde{J}_c(\mu_{\rm eff},k)
=\sum_{\l=1}^\infty\widetilde{c}_\l(k)\re^{-2\l\mu_{\rm eff}}. 
\end{align}
Of course, these coefficients are completely determined by the original coefficients $a_\l(k)$, $b_\l(k)$ and $c_\l(k)$. It was conjectured in \cite{hmo2} that one has the following relationship, 
\begin{align}
\widetilde{c}_\l(k)=- k^2 \frac{\partial}{\partial k} 
\left(\frac{\widetilde{b}_\l(k)}{2\l k}\right).
\label{bcrel}
\end{align}
This means that there are only two sets of independent coefficients left in (\ref{gen-M2}), which we will take to be $a_\ell (k)$, $b_\ell (k)$. The calculation of the 
non-perturbative grand potential is then reduced to the determination of these two sets of coefficients. We will now review the techniques to calculate these coefficients, as well as the evidence for the above conjectures. 

\subsection{Calculating the grand potential}

The grand potential of the Fermi gas can be in principle computed from the one-particle Hamiltonian (\ref{onepH}). However, this Hamiltonian is not exactly solvable and in order to find analytic answers one needs some sort of approximation. One obvious possibility is to use the WKB approximation. Since $k$ plays the role of the Planck constant, the WKB approximation leads to perturbative expansions around $k=0$. For example, the coefficient $a_\ell(k)$ has an expansion of the form, 
\be
\label{a-WKB}
a_\ell(k)= {1\over k} \sum_{n=0}^\infty a_{\ell,n} k^{2n}. 
\ee
A similar expansion holds for $b_\ell(k)$ and $c_\ell(k)$, with coefficients $b_{\ell,n}$, $c_{\ell,n}$, respectively. The coefficients of these expansions can be in principle calculated systematically, and this provides a valuable source of information about the contribution of membrane instantons. Unfortunately, in the WKB approximation, the calculation of worldsheet instantons and bound states is difficult since these effects are non-perturbative in $k$. 

An important technical and conceptual tool in the analysis of the Fermi gas is that, as noticed in \cite{mp-fermi}, its grand potential can be computed by using a variant of the Thermodynamic Bethe Ansatz (TBA). This formulation is based on \cite{zamo,tw} and it has been explored in \cite{hmo,py,hmo1,cm,hmo2}. Let us briefly review it here. Consider the set of coupled non-linear equations for the two functions $\epsilon(x)$, $\eta(x)$:
\be
\label{tba-k}
\ba
U(x)&= \epsilon\left(x\right) + \int_{-\infty}^{\infty} {\rd x' \over 2\pi k } {\log \left(1+\eta^2(x')\right) \over \cosh\left({x-x' \over k} \right)}, \\
\eta(x)&=-z \int_{-\infty}^{\infty} {\rd x'  \over  2\pi k } {\re^{-\epsilon(x')}\over \cosh\left({x-x' \over k} \right)},
\ea
\ee
where 
\be
\label{ux}
U(x)= \log \left( 2 \cosh {x\over 2} \right).
\ee
Let us also define, 
\be
\ba
R_+(x|z)&= \re^{-\epsilon(x)},\\
R_-(x|z)&=R_+(x |z) \int_{-\infty}^{\infty} {\rd x'  \over \pi k} \, {\arctan\, \eta(x') \over   \cosh^2\left({x - x'\over k} \right)},
\ea
\ee
which are even and odd functions of $z$, respectively. Let us denote the even and odd parts of $J(z)$ as,
\be
\label{eoj}
J_\pm (z)={1\over 2} \left( J(z) \pm J(-z) \right).
\ee
Then, one has
\begin{align}
{\partial J_\pm \over \partial z}= {1\over 4 \pi k} \int_{-\infty}^{\infty} \rd x R_\pm\left( x |z\right), 
\end{align}
and this makes it possible to calculate the grand potential from the solution to the TBA equations (\ref{tba-k}). 

The TBA equations can be analyzed in two different ways. First, one can study them in the small $k$ regime, but exactly in $\mu$. This is equivalent to the WKB expansion of the Fermi gas. It leads to a very efficient method to calculate the expansion around $k=0$ of the membrane instanton contribution \cite{cm}. Second, one can study them at fixed values of $k$, and for small $\mu$. This makes it possible to calculate the partition function $Z(N,k)$ for small values of $N$ \cite{hmo,py}. Notice that, in order to 
make contact with the expansion (\ref{gen-M2}), one should study the TBA equations at large $\mu$ but exactly in $k$. 
However, as pointed out in \cite{cm}, one encounters a phase transition as $\mu$ grows (related to Bose--Einstein condensation) 
which makes it difficult to obtain analytic results in this regime. In \cite{hmo1}, the values of $Z(N,k)$ for small $N$ 
were extrapolated numerically to obtain precise estimates of the first terms appearing in the large $\mu$ expansion of the grand potential. 
One finds, for example, for $k=1,2$ \cite{hmo1}, 
\be
\ba
J^{\rm (np)}(1,\mu)&=\biggl[\frac{4\mu^2+\mu+1/4}{\pi^2}\biggr]\re^{-4\mu}
+\biggl[-\frac{52\mu^2+\mu/2+9/16}{2\pi^2}+2\biggr]\re^{-8\mu}\\
&\quad+\biggl[\frac{736\mu^2-152\mu/3+77/18}{3\pi^2}-32\biggr]\re^{-12\mu}+\CO(\re^{-16\mu}), \\\
J^{\rm (np)}(2,\mu)&=\biggl[\frac{4\mu^2+2\mu+1}{\pi^2}\biggr]\re^{-2\mu}
+\biggl[-\frac{52\mu^2+\mu+9/4}{2\pi^2}+2\biggr]\re^{-4\mu}
\\
&\quad+\biggl[\frac{736\mu^2-304\mu/3+154/9}{3\pi^2}-32\biggr]\re^{-6\mu} +\CO(\re^{-8\mu}). 
\ea
\ee

Finally, a very important guiding principle in the determination of the grand potential is the cancellation mechanism discovered in \cite{hmo1}, which we will call, following \cite{cm}, the HMO cancellation mechanism. This mechanism is based on the following observation: the Gopakumar--Vafa representation (\ref{gvone}) 
of $J^{\rm WS}(\mu, k)$ shows that it has 
double poles at all integer values of $k$. Since the original matrix integral (\ref{abjmmatrix}) is not singular for any value of $k$, 
there must be some way of canceling these divergences. The proposal in \cite{hmo1} is that, for any fixed, integer value of $k$, singularities are cancelled 
order by order in the expansion in ${\rm e}^{-\mu}$. Generically there are many bound states $(\ell, m)$ which contribute  to a given order in ${\rm e}^{-\mu}$, therefore the HMO cancellation mechanism gives a precise relationship among the pole structure of these contributions \cite{hmo2}. 

The HMO mechanism has deep conceptual implications in M-theory. It shows, in a precise and quantitative way, that the genus expansion based on 
perturbative strings is essentially meaningless: in the non-perturbative completion of type IIA string theory at finite, integer $k$ through M-theory, only the combination of membrane instantons, worldsheet instantons and their bound states makes sense. In practice, this mechanism can be used to constrain the structure of the contribution of bound states, and this leads to the conjecture (\ref{conj-one}) of \cite{hmo2}. In combination with the WKB expansion, it also leads to closed, conjectural expressions for the first few coefficients $a_\ell(k)$, $b_\ell(k)$ and $c_\ell(k)$ \cite{hmo1,cm,hmo2}. One finds, for example, for the first $a_\ell$, up to $\ell=3$, 
\be
\label{a1a2a3}
\ba
a_1(k)&=-\frac{4}{\pi^2 k} \cos \Bigl(\frac{\pi k}{2} \Bigr),\\
a_2(k)&=-\frac{2}{\pi^2 k}\bigl(4+5\cos(\pi k)\bigr),\\
a_3(k)&=-\frac{8}{3\pi^2k}\cos\Bigl(\frac{\pi k}{2}\Bigr)
\bigl(19+28\cos(\pi k)+3\cos(2\pi k)\bigr),
\ea
\ee
and for the $b_\ell(k)$, also up to $\ell=3$, 
\be
\label{b-bare}
\ba
b_1(k)&=  \frac{2}{\pi}\cos^2\left(\frac{\pi k}{2}\right)\csc\left(\frac{\pi k}{2}\right), \\
b_2(k)&={4 \over \pi^2 k} \left( 1 + \cos\left( \pi k\right)  \right) + {1\over 2 \pi }\csc \left( \pi k\right) \left( 17 + 24 \cos\left( \pi k\right) + 9 
\cos\left(2 \pi k\right)  \right), \\
b_3(k)&=\frac{4}{\pi^2 k}
\left[13\cos\(\frac{\pi k}{2}\)+5\cos\(\frac{3\pi k}{2}\)\right]\\
&\hspace{0.5cm}+\frac{1}{3\pi}\csc\(\frac{3\pi k}{2}\)
(241+405\cos(\pi k)+222\cos(2\pi k)+79\cos(3\pi k)+9\cos(4\pi k)).
\ea
\ee
From these values, one can deduce the $\widetilde{b}_\ell(k)$, which have a somewhat simpler expression, 
\be
\label{b1b2b3}
\ba
\widetilde{b}_1(k)&=\frac{2}{\pi}\cot\(\frac{\pi k}{2}\)
\cos\(\frac{\pi k}{2}\),\\
\widetilde{b}_2(k)&=\frac{1}{\pi}\cot(\pi k)\bigl(4+5\cos(\pi k)\bigr),\\
\widetilde{b}_3(k)&=\frac{4}{3\pi}
\cot\(\frac{3\pi k}{2}\)
\cos\(\frac{\pi k}{2}\)
\bigl(13+19\cos(\pi k)+9\cos(2\pi k)\bigr).
\ea
\ee

In summary, the results of \cite{mp-fermi,hmo,py,hmo1,cm,hmo2} lead to concrete results as well as precise conjectures on the grand potential $J(\mu)$. The contribution of worldsheet instantons is fully determined by topological string theory on local $\IP^1 \times \IP^1$. The conjectures (\ref{conj-one}) and (\ref{bcrel}) reduce the calculation of membrane instanton and bound state contributions to the determination of the coefficients $a_\ell(k)$, $b_\ell(k)$. These conjectures have been tested in detail against the WKB expansions of the TBA/Fermi gas, numerical extrapolations at large $N$, and the HMO cancellation mechanism. Moreover, using all these inputs, it is possible to conjecture the exact form of the very first coefficients $a_\ell(k)$, $b_\ell(k)$.

\section{Membrane instantons as quantum periods}

Clearly, in order to give a complete description of $J^{\rm np}(\mu)$, we need an efficient way to compute the coefficients 
$a_\ell(k)$, $b_\ell(k)$ exactly as a function of $k$. The goal of this section is to provide overwhelming evidence that these coefficients are 
determined by the refined topological string on local $\IP^1 \times \IP^1$ in the NS limit, 
and therefore \cite{mm1, mm2,acdkv} by the quantum periods of the spectral curve of local $\IP^1 \times \IP^1$. 
This fact allows us to compute these coefficients systematically up to any desired order.
We first observe that the leading order coefficients $a_{\ell,0}$ and $b_{\ell,0}$ in the WKB expansion \eqref{a-WKB}
are interpreted as classical periods in the topological string.
We then find that the full coefficients $a_\ell(k)$ and $b_\ell(k)$ just correspond to the ``quantum'' corrected periods.
We also show that these expressions indeed guarantee the pole cancellation coming from the worldsheet instanton correction.

\subsection{The refined topological string}
\label{ref-rev}
When defined on a local CY manifold, topological string theory can be ``refined" by introducing a further coupling constant. This refinement has its origin in Nekrasov's instanton calculation 
in $\CN=2$ gauge theories, and it can be interpreted in terms of a so-called 
``Omega background" for the gauge theory \cite{nekrasov}. In topological string theory, a natural point of view to understand 
the refinement is to consider the GV invariants which appear in the large radius expansion (\ref{gv-exp}) of the topological string free energy \cite{ref-vertex}. These invariants 
are interpreted by considering M-theory compactified on the CY $X$. In this compactification, M2 branes 
wrapping a two-cycle of $X$ with degree ${\bf d}$ lead to BPS states in five dimensions, 
with spins $(j_L, j_R)$ with respect to the rotation group $SU(2)_L \times SU(2)_R$. The index for such states, 
which we denote by $N^{\bf d}_{j_L, j_R}$, is not invariant under deformations on a general CY manifold. 
However, in a {\it local} CY it is a topological invariant and it can be used to define the refined topological string free energy 
at large radius. In order to write down this free energy, we introduce the $SU(2)$ character, 
\be
\chi_{j}(q)
={q^{2j+1}-q^{-2j-1} \over q-q^{-1}}. 
\ee
Notice that, with our conventions, $j_L$ and $j_R$ are generically half-integers.  
The refined topological string free energy is a function of the K\"ahler moduli $T_I$, $I=1, \cdots, n$, and two 
parameters $\epsilon_{1,2}$ which ``refine" the topological string coupling constant. We also introduce (see for example \cite{hk,ckk}) 
\be
\epsilon_L={\epsilon_1 -\epsilon_2 \over 2},\qquad
\epsilon_R= {\epsilon_1 +\epsilon_2 \over 2},
\ee
and
\be
q_{1,2}=\re^{\epsilon_{1,2}},  \qquad q_{L,R}= \re^{\epsilon_{L,R}}.
\ee
Then, the refined topological string free energy, in terms of the BPS index $N^{\bf d}_{j_L, j_R}$, is given by 
\be
\label{refBPS}
F({\bf Q}, \epsilon_1, \epsilon_2)=
\sum_{j_L, j_R \ge 0 }\sum_{w\ge 1}\sum_{{\bf d}} 
{1\over w}  N_{j_L, j_R}^{\bf d}  {\chi_{j_L} (q^w_L) \chi_{j_R} (q^w_R) \over \left(q_1^{w/2} - q_1^{-w/2}\right) \left(q_2^{w/2} -q_2^{-w/2} \right)}
{\bf Q}^{w{\bf d}}. 
\ee
It is also very useful to introduce another set of integer invariants $n_{g_L, g_R}^{\bf d}$ by the following equality of generating functionals, 
\be
\label{change-basis}
\ba
&\sum_{j_L, j_R\ge 0}  N_{j_L, j_R}^{\bf d}
\chi_{j_L}(q_L) \chi_{j_R}(q_R)\\
&=\sum_{g_L, g_R \ge 0 } (-1)^{g_L+g_R} n_{g_L, g_R}^{\bf d}
\left(q_L^{1 /2} - q_L^{-1 /2} \right)^{2g_L}
\left(q_R^{1 /2} - q_R^{-1 /2} \right)^{2g_R}.
\ea
\ee
In terms of these invariants, the refined free energy reads
\be
\label{ref-free-2}
F({\bf Q}, \epsilon_1, \epsilon_2)=
\sum_{g_L, g_R \ge 0 }\sum_{w\ge 1}\sum_{{\bf d}} 
{1\over w} (-1)^{g_L+g_R} n_{g_L, g_R}^{\bf d}  
{ \left(q_L^{w /2} - q_L^{-w /2} \right)^{2g_L} \over q_1^{w/2} - q_1^{-w/2} }
{ \left(q_R^{w /2} - q_R^{-w /2} \right)^{2g_R} \over q_2^{w/2} -q_2^{-w/2} }
{\bf Q}^{w{\bf d}}.
\ee
Sometimes it is also useful to consider the perturbative expansion of the free energy, which following \cite{hk} we will write as
\be
\label{ref-pert}
 F({\bf Q}, \epsilon_1, \epsilon_2)= \sum_{n,g\ge 0} \left( \epsilon_1+ \epsilon_2 \right)^{2n} \left( \epsilon_1 \epsilon_2\right)^{g-1} F^{(n,g)}( {\bf Q}).
 \ee
The standard topological string is a particular case of the refined topological string, corresponding to 
\be
\epsilon_1=-\epsilon_2=g_s. 
\ee
In this limit, since $q_R=1$, the only invariants which survive in (\ref{ref-free-2}) have $g_R=0$. They correspond to the original 
Gopakumar--Vafa invariants appearing in (\ref{gv-exp}), 
\be
n_g^{\bf d}=n_{g,0}^{\bf d}, 
\ee
and the expression (\ref{ref-free-2}) becomes the original Gopakumar--Vafa formula for the topological string free energy (\ref{gv-exp}). 
In the expansion (\ref{ref-pert}), only the terms with $n=0$ 
survive, and one recovers the genus expansion of standard topological string theory. In particular, 
\be
F^{(0,g)}({\bf Q}) = (-1)^{g-1} F_g({\bf Q})
\ee
is, up to a sign, the genus $g$ free energy of topological strings. 

One obvious problem in topological string theory is the calculation of the refined free energy and of the corresponding BPS invariants. At large radius, one can 
use the refined topological vertex of \cite{ref-vertex}, or formulate the problem in terms of stable pair invariants \cite{ckk}. From the B-model point of view, 
a refined version of the holomorphic anomaly equation of \cite{bcov} has been conjectured in \cite{kw,hk}. In this formulation one can calculate the 
refined string free energy in any 
symplectic frame (appropriate for example for orbifold and conifold points).  

There is a special limit of the refined topological string which was first identified in \cite{ns} and has remarkable properties. In this limit, 
one of the parameters $(\epsilon_1,
\epsilon_2)$ goes to zero while the other is kept finite,
\be
 \epsilon_1=\hbar, \qquad \epsilon_2 \to 0.
\ee
This is usually called the Nekrasov--Shatashvili (NS) limit. The refined free energy (\ref{ref-free-2}) has a simple pole in this limit, and we define the NS free energy as
\be
F_{\rm NS}({\bf Q}, \hbar) = \lim_{\epsilon_2 \rightarrow 0} 
\epsilon_2 F({\bf Q}, \epsilon_1, \epsilon_2), 
\ee
which has the following expansion at large radius in terms of integer invariants,
\be
\label{NS-GV}
F_{\rm NS}({\bf Q}, \hbar) =\sum_{g=0}^\infty \sum_{w\ge 1} \sum_{{\bf d}} 
{1\over w^2}\hat n_{g}^{\bf d}  {  \left(q ^{w /4} - q^{-w /4} \right)^{2g } 
\over q^{w/2} -q^{-w/2} } {\bf Q}^{w{\bf d}},
\ee
where
\be
\label{qh}
q=\re^{\hbar}
\ee
and 
\be
\label{ns-invs}
\hat n_g^{{\bf d}}= \sum_{g_L+g_R=g} (-1)^g n_{g_L, g_R}^{{\bf d}}
\ee
are the integer invariants appearing in the NS limit. They shouldn't be confused with the original Gopakumar--Vafa invariants appearing in (\ref{gv-exp}). An expression equivalent to (\ref{NS-GV}), which will be useful later on, is 
\be
\label{NS-j}
F_{\rm NS}({\bf Q}, \hbar) =\sum_{j_L, j_R \ge 0} \sum_{w\ge 1} \sum_{{\bf d}} 
{1\over w^2}N^{\bf d}_{j_L, j_R} {  \chi_{j_L} (q^{w/2}) \chi_{j_R} (q^{w/2}) 
\over q^{w/2} -q^{-w/2} } {\bf Q}^{w{\bf d}}.
\ee
From (\ref{ref-pert}) one finds the following 
perturbative expansion
\be
F_{\rm NS}({\bf Q}, \hbar)=\sum_{n \ge 0} \hbar^{2n-1} F^{(n,0)}({\bf Q}),
\ee
and the first term corresponds (up to an overall sign) to the genus zero free energy, i.e. to the classical prepotential of special geometry. The free energies $F^{(n,0)}$ have been studied from the point of view of the holomorphic 
anomaly equations in \cite{huang}. 

It is well known that, in some cases, $\CN=2$ supersymmetric gauge theories can be 
engineered as particular limits of topological string theory \cite{kkv-ge}. In these cases, 
the topological string free energy in the NS limit has been related \cite{ns} to the quantization of the classical integrable system associated to the $\CN=2$ theory. In \cite{mm1,mm2}, Mironov and Morozov interpreted the 
quantization in terms of quantum periods of the spectral curve describing the integrable system. The appearance of quantum periods was clarified and extended to general, local CY geometries in \cite{acdkv} from the point of view of mirror symmetry. In the local B-model, the mirror CY geometries are described by a curve of the form 
\be
\label{spec-curve}
H(x,p)=0. 
\ee
This describes a genus $n$ Riemann surface and defines (locally) a function $p(x)$. Let us choose a symplectic basis 
$A_I$, $B_I$, $I=1, \cdots, n$. The classical periods of the meromorphic one-form 
\be
\lambda= p(x) \rd x,
\ee
are given by
\be
\Pi_{A_I}(z_I)=\oint_{A_I} \lambda, \qquad \Pi_{B_I}(z_I)=\oint_{B_I} \lambda, \qquad I=1, \cdots, n. 
\ee
Here, the $z_I$ are complex deformation parameters appearing in the equation of the spectral curve (\ref{spec-curve}). 
In terms of these classical periods, the classical prepotential $F^{(0,0)}$ is defined 
as follows:
\be
-T_I= \Pi_{A_I}(z_I), \qquad -Q_I \partial_{Q_I}F^{(0,0)}= \Pi_{B_I}(z_I), \qquad I=1, \cdots, n. 
\label{class-pers}
\ee
The first equation gives the mirror map, relating the flat coordinates $T_I$ to the complex deformation parameters $z_I$. 
We can now ``quantize" the classical spectral curve (\ref{spec-curve}) by promoting $x$, $p$ to operators $\hat x$, $\hat p$ with commutation relations
\be
[\hat x, \hat p]=-\hbar, 
\ee
so that, in position space, $\hat p$ acts as $\hbar \partial_x$. The quantization of the spectral curve amounts to solving the time-independent Schr\"odinger equation 
\be
\label{schr}
H \left( x, \hbar \partial_x\right) \Psi(x, \hbar)=0. 
\ee
We then write
\be
\Psi(x, \hbar)= \exp \left({1\over \hbar} S(x, \hbar)  \right) 
\ee
and interpret 
\be
\partial S= \partial_x S(x, \hbar) \rd x
\ee
as a ``quantum" differential. Indeed, the function $S(x,\hbar)$ has a WKB expansion 
\be
S(x, \hbar)= \sum_{n\ge 0}  S_n (x) \hbar^{2n},
\ee
and it follows immediately from (\ref{schr}) that 
\be
\partial_x S_0(x)= p(x).
\ee
One can then consider the quantum periods
\be
\label{q-pers-def}
\Pi_{A_I}(z_I;\hbar)= \oint_{A_I} \partial S, \qquad \Pi_{B_I}(z_I;\hbar)= \oint_{B_I} \partial S, \qquad I=1, \cdots, n, 
\ee
which define the ``quantum" mirror map and quantum prepotential $F_{\rm NS}(\hbar)$ through, 
\be
\label{q-pers}
-T_I(\hbar)= \Pi_{A_I}(z_I;\hbar), \qquad -Q_I \partial_{Q_I}F_{\rm NS}(\hbar)={1\over \hbar} \Pi_{B_I}(z_I;\hbar), \qquad I=1, \cdots, n. 
\ee
 It is clear from the above discussion that, in the WKB expansion, 
\be
F_{\rm NS}(\hbar)= {1\over \hbar} F^{(0,0)}+\CO (\hbar^2)
\ee
The claim of \cite{mm1,mm2,acdkv}, following \cite{ns}, is that the function $F_{\rm NS}(\hbar)$ defined 
in this way is the NS limit of the refined free energy of topological string theory. 
This claim was tested in various examples by comparing the result obtained from the quantum periods with 
existing results on $F_{\rm NS}(\hbar)$. In \cite{acdkv} the approach based on quantum periods 
was justified by using the dual matrix model description of the refined string in terms of $\beta$-ensembles, which exists in some cases, and deriving the 
Schr\"odinger equation directly in the matrix model (such a derivation was first discussed in \cite{bt}).

\subsection{Classical limit}\label{sec:cl}

Our goal in this section is to show that the functions $J_a(\mu, k)$, $J_b(\mu, k)$ are essentially quantum periods of the local $\IP^1 \times \IP^1$ geometry, where 
$k$ plays the r\^ole of the quantum deformation parameter $\hbar$. Since quantum periods become the classical periods of special geometry when $\hbar \to 0$, 
we will first show that the classical limits of $J_{a,b}(\mu, k)$, which can be computed in closed form, are these classical periods. 

Let us then consider the leading coefficients $a_{\ell,0}$ and $b_{\ell,0}$ in the WKB expansion, defined in (\ref{a-WKB}) (and a similar equation for $b_\ell(k)$).
We first notice that in the WKB expansion of the even/odd parts of the grand potential, 
\be
J_\pm(\mu,k)=\frac{1}{k}\sum_{n=0}^\infty J_{\pm,n}(\mu) k^{2n},
\ee
the leading order corrections are given by \cite{mp-fermi,cm}
\be
\frac{\partial J_{+,0}}{\partial z}=\frac{1}{2\pi} K\left( \frac{z^2}{16}\right),\qquad
\frac{\partial J_{-,0}}{\partial z}=-\frac{z}{4\pi^2}\,_3F_2\left( 1,1,1;\frac{3}{2},\frac{3}{2};\frac{z^2}{16}\right).
\label{eq:J0}
\ee
As observed in \cite{cm}, the coefficients $a_{\ell,0}$ and $b_{\ell,0}$ can be read off only from $\partial_z J_{+,0}$,
and we have the relation
\be
\frac{\partial J_{+,0}^{\rm M2}}{\partial z}=
\sum_{\ell\ge 1}  \left[  \left( 2 \pi \ri \log z - \pi^2 \right)  \ell a_{\ell,0} + \pi \ri \left(\ell b_{\ell,0} - a_{\ell,0} \right)  \right] z^{-2\ell-1}.
\ee
Using the asymptotic expansion of \eqref{eq:J0}, one finds,
\be
\ba
a_{\ell,0}&= -{1\over  \pi^2 \ell} \left( {\Gamma \left( \ell+{1\over 2} \right) \over 
\Gamma({1\over 2}) \ell!} \right)^2 16^\ell, \\
b_{\ell,0}&= {2 \over \pi^2 \ell} \left( {\Gamma \left(  \ell+{1\over 2} \right) \over 
\Gamma({1\over 2})  \ell!} \right)^2 16^ \ell \left[ \psi\left( \ell+{1\over 2} \right) -\psi( \ell+1)+ 2 \log 2-{1\over 2 \ell} \right],
\ea
\ee
where $\psi(x)$ is the digamma function.

Now, we would like to show that the functions
\be
J_{a,0}(\mu)=\sum_{\ell=1}^\infty a_{\ell,0}e^{-2\ell \mu},\qquad
J_{b,0}(\mu)=\sum_{\ell=1}^\infty b_{\ell,0}e^{-2\ell \mu},
\ee
are related to the classical periods (\ref{class-pers}) of the topological string on local $\IP^1 \times \IP^1$, restricted to the ``diagonal" case $T_1=T_2$. 
A useful approach to determine the periods is to use the fact that they solve 
differential equations of the Picard--Fuchs type. In the case of local $\IP^1 \times \IP^1$, the periods are in general functions of two moduli for deformations of the complex 
structure, $z_1$ and $z_2$. They are annihilated 
by the pair of differential operators (see for example \cite{hkr,dmp} for a summary of these results)
\be
\label{pfsystem}
\ba
{\cal L}_1 &= z_2 (1-4 z_2) \xi^2_2 - 4 z_1^2 \xi_1^2 -
8 z_1 z_2 \xi_1 \xi_2 - 6 z_1\xi_1 +(1- 6 z_2) \xi_2, \\
{\cal L}_2 &= z_1 (1-4 z_1) \xi^2_1 - 4 z_2^2 \xi_2^2 -
8 z_1 z_2 \xi_1 \xi_2 - 6 z_2\xi_2 +  (1-6 z_1)\xi_1, 
\ea
\ee
where 
\be
\xi_i={\partial\over \partial z_i}.
\ee
The $A$-periods are given by 
\be
\Pi_{A_I}(z) =\log z_I + \widetilde \Pi_A (z_1, z_2), \qquad I=1,2,
\ee
where
\be
\label{Aper}
\widetilde \Pi_A (z_1, z_2)= 2\sum_{k,l\ge 0, \atop (k,l)\not=(0,0)} { \Gamma(2k + 2l) \over \Gamma(1+k)^2 \Gamma(1+l)^2} z_1^k z_2^l =2z_1 + 2z_2 + 3 z_1^2 + 12 z_1 z_2 + 3 z_2^2 + \cdots
\ee
There are two independent $B$-periods, $\Pi_{B_I}(z_1, z_2)$, $I=1,2$, which are related by the exchange of $z_1$ and $z_2$, 
\be
\Pi_{B_2}(z_1, z_2)=\Pi_{B_1}(z_2, z_1). 
\ee
The $B_1$ period is given by 
\be
\label{B1per}
\Pi_{B_1}(z_1, z_2)=-{1\over 8}\left( \log^2 z_1  -2 \log z_1\log z_2 -\log^2 z_2 \right) +{1\over 2} \log z_2\,  \widetilde \Pi_A (z_1, z_2)+ {1\over 4}  \widetilde \Pi_B (z_1, z_2) 
\ee
where
\be
\label{Bper}
\ba
\widetilde \Pi_B (z_1, z_2) &= 8 \sum_{k,l\ge 0, \atop (k,l)\not=(0,0)} { \Gamma(2k + 2l) \over \Gamma(1+k)^2 \Gamma(1+l)^2} \left( \psi(2k+2l) -\psi(1+l)\right) z_1^k z_2^l \\
 &= 8 z_1+ 22 z_1^2 + 40 z_1 z_2 + 4z_2^2 +\cdots
 \ea
 \ee
 Notice that the diagonal case $T_1=T_2 $ simply corresponds to $z_1=z_2=z$. One finds, after this specialization, that
 \be
\ba
\label{id-0}
J_{a,0}(\mu)&=-\frac{1}{\pi^2} \widetilde{\Pi}_{A}(z,z),\\
J_{b,0}(\mu)&=\frac{1}{2\pi^2} \widetilde{\Pi}_{B}(z,z),
\ea
\ee
under the identification\footnote{
One should not confuse this deformation parameter $z=\re^{-2\mu}$ in the topological string 
with the fugacity $z=\re^\mu$ in the Fermi-gas system.} 
\be
\label{z-mu}
z=\re^{-2\mu}.
\ee

The relation (\ref{id-0}) can be proved by resumming the coefficients in (\ref{Aper}) and (\ref{Bper}). Equivalently, one can restrict the problem to the one-modulus case from the very beginning. 
The Picard--Fuchs operator for local $\IP^1 \times \IP^1$ along the diagonal $z_1=z_1=z$ takes the form \cite{kz}
\be
\label{pf}
\CL= \theta^3-16 z \prod_{i=1}^3 \left( \theta- a_i+1 \right), 
\ee
where
\be
\theta =z {\rd \over \rd z}
\ee
and the constants $a_i$ $(i=1,2,3)$ are given by
\be
a_1={1\over 2}, \quad a_2={1\over 2}, \quad a_3=1.
\ee
The A- and B-periods can then be found by using the Frobenius method: 
one first computes the fundamental period,
\be
\label{fund_period}
\Pi_0(z, \rho)= \sum_{n\ge 0} a_n(\rho) z^{n+\rho},
\ee
with
\be
\label{fund_period_sol}
a_n(\rho)=16^n { \Gamma^2\left( n+ \rho+{1\over 2} \right) \Gamma(n+\rho) \over \Gamma^3( n+\rho+1)} {\Gamma^3(\rho+1) \over \Gamma^2(\rho+{1\over 2}) \Gamma(\rho)}.
\ee
The A- and B-periods are then given by
\be
\label{period-cl}
\Pi_{A}(z,z)=\frac{\rd\varpi_0(z,\rho)}{\rd\rho}\biggr|_{\rho=0},\qquad
\Pi_{B}(z,z)={1\over 4} \frac{\rd^2\varpi_0(z,\rho)}{\rd\rho^2}\biggr|_{\rho=0}.
\ee
In this way one finds, 
\be
\ba
\widetilde \Pi_{A}(z,z)&=\sum_{n \ge 0}  {1\over  n} \left( {\Gamma \left( n+{1\over 2} \right) \over 
\Gamma({1\over 2}) n!} \right)^2 16^n z^n, \\
\widetilde \Pi_{B}(z,z)&=\sum_{n \ge 0}   {4 \over  n} \left( {\Gamma \left( n+{1\over 2} \right) \over 
\Gamma({1\over 2}) n!} \right)^2 16^n \left[ \psi\left(n+{1\over 2} \right) -\psi(n+1)+ 2 \log 2-{1\over 2n} \right] z^n, 
\ea
\ee
and one verifies (\ref{id-0}). 

Thus we conclude that the leading order membrane instanton corrections
are interpreted as the classical periods in the topological string on
local $\IP^1 \times \IP^1$.
To avoid confusion, note that the notation of $a_\ell(k),b_\ell(k)$
comes from the Fermi gas formalism of ABJM theory, while the A- and
B-periods are the standard notation in special geometry.
Surprisingly, these two notations match well.

\subsection{Membrane instantons and quantum periods}

In the previous subsection, we have observed that the leading order
functions $J_{a,0}(\mu)$ and $J_{b,0}(\mu)$ correspond to the
classical periods of the topological string on local
$\IP^1 \times\IP^1$. Let us recall that the Chern-Simons level $k$ plays the role of the
Planck constant in the Fermi gas formulation, see (\ref{planck-fermi}).
This suggests that the counterparts of the full functions
$J_{a}(\mu,k)$ and $J_{b}(\mu,k)$ may be certain ``quantum'' corrected
periods. This is indeed the case: in this subsection, we will give overwhelming evidence that the functions $J_{a}(\mu,k)$ and
$J_{b}(\mu,k)$ correspond to the quantum periods, in the sense of
\cite{mm1,mm2,acdkv} reviewed above. Notice however that the quantum parameter $\hbar$ is not the 
topological string coupling constant, but rather its inverse. Therefore, the weakly coupled WKB expansion of the 
quantum periods corresponds here, as in the Fermi gas approach, to a strongly coupled 
topological string. 

Our evidence is based on direct computation of the quantum periods and comparison to the known results of quantum grand potential. 
Let us first consider the A-period.
The quantum correction to the classical A-period was computed in \cite{acdkv}.
We briefly  review their method here.
We start with the mirror curve for local $\IP^1 \times \IP^1$,
\be
-1+\re^x+\re^p+z_1\re^{-x}+z_2\re^{-p}=0.
\ee
In this curve, $z_{1,2}$ are the complex deformation parameters of the geometry appearing in (\ref{pfsystem}). 
The Schr\"odinger equation (\ref{schr}) reads in this case, 
\be
(-1+\re^x+z_1 \re^{-x})\Psi(x)+\Psi(x+\hbar)+z_2 \Psi(x-\hbar)=0.
\ee
This equation can be solved perturbatively in $\hbar$, by using the WKB expansion. 
However, there is a more efficient way to compute the periods, at all orders in $\hbar$ \cite{acdkv}.\footnote{We are thankful to D. Krefl for detailed explanations on this issue.}
For this purpose, we introduce the new function
\be
\label{psi-psi}
V(x)=\frac{\Psi(x+\hbar)}{\Psi(x)}.
\ee
Then, we obtain the difference equation
\be
\label{eq:diff_eq}
V(x)=1-\re^x+z_1\re^{-x}+\frac{z_2}{V(x-\hbar)}.
\ee
This equation can be solved around $z_1=z_2=0$ in a power series expansion,
\be
V(X)=1-X+\frac{z_1}{X}+\frac{z_2}{1-q^{-1}X}+{\cal O}(z_i^2),
\ee
where $X=\re^x$ and $q=\re^\hbar$, as in (\ref{qh}). The quantum A-period is given by 
\be
\Pi_{A_I}(z_1,z_2;q)=\log z_I +\widetilde{\Pi}_{A}(z_1,z_2;q) \qquad I=1,2,
\ee
where $\widetilde{\Pi}_{A}$ is given by 
\be
\widetilde{\Pi}_{A}(z_1,z_2;q)= -2 \, {\rm Res}_{X=0} \log \left[V(X) \over 1-X \right].
\ee
Notice that here we write the dependence on the quantum parameter $\hbar$ in exponentiated form, through $q$. 
This quantum period defines the quantum mirror map,
\be
\label{qmm}
Q_I (z_1, z_2; q)= z_I \re^{\widetilde{\Pi}_{A}(z_1,z_2;q)}, \qquad I=1,2.
\ee
Using the solution of \eqref{eq:diff_eq}, we find
\be
\label{q-Aper}
\ba
\widetilde{\Pi}_{A}(z_1,z_2;q)&=2(z_1+z_2)+3(z_1^2+z_2^2)+2(4+q+q^{-1})z_1z_2+\frac{20}{3}(z_1^3+z_2^3) \\
&\quad +2(16+6q+6q^{-1}+q^2+q^{-2})z_1 z_2(z_1+z_2)+{\cal O}(z_i^4).
\ea
\ee
where $q$ is given in (\ref{qh}). In the classical limit $q\rightarrow 1$, the quantum period becomes the classical period (\ref{Aper}):
\be
\lim_{q\to 1} \widetilde{\Pi}_{A}(z_1,z_2;q)=\widetilde{\Pi}_{A}(z_1,z_2), 
\ee
as it should. 

Let us now make contact with the function $J_a(\mu, k)$ appearing in the membrane instanton contribution to ABJM theory. 
If we identify the parameters $z_1,z_2$ as 
\be
\label{eq:z1z2-z}
z_1=q^{\frac{1}{2}}z,\qquad z_2=q^{-\frac{1}{2}}z,
\ee
the quantum A-period becomes
\be
\ba
\widetilde{\Pi}_{A}(q^{\frac{1}{2}}z,q^{-\frac{1}{2}}z;q)&= 2(q^{1/2} + q^{-1/2}) z + \left( 8 + 5( q^{1/2} + q^{-1/2}) \right) z^2\\
& + 
\frac{2 \left(3 q^5+31 q^4+66 q^3+66 q^2+31 q+3\right)}{3 q^{5/2}} z^3+ \CO(z^4)
\ea
\ee
If we now set the deformation parameter $\hbar$ to be
\be
\label{eq:q-k}
\hbar=\pi \ri k, 
\ee
or, equivalently, 
\be
\label{q-k}
q=\re^{\pi \ri k}, 
\ee
we see that the coefficient of $z^\ell$ in the quantum period just
gives the membrane instanton coefficients $a_\ell(k)$ appearing in \eqref{a1a2a3}, up to an overall factor of $-1/\pi^2$. In fact, we have checked that 
this is the case up to order $\ell=10$ by comparing the quantum period to the result for the coefficients $a_\ell(k)$ obtained from the TBA equations and explained in Appendix A. 

We conclude that the function $J_a(\mu,k)$ is given, up to an overall constant, by the
quantum A-period evaluated on the slice (\ref{eq:z1z2-z}) ,
\be
\label{A-ja}
J_a(\mu,k)=-\frac{1}{\pi^2 k} \widetilde{\Pi}_A\left(q^{\frac{1}{2}}z,q^{-\frac{1}{2}}z;q\right),
\ee
and with the identifications (\ref{eq:q-k}) and (\ref{z-mu}). 

Let us now consider the quantum B-periods $\Pi_{B_I}(z_1,z_2;q)$. As in the undeformed, classical case, there are two of them, 
but they are related by the exchange of the moduli, 
\be
\Pi_{B_2}(z_1,z_2;q)=\Pi_{B_1}(z_2,z_1;q).
\ee
The quantum counterpart of (\ref{B1per}) is 
\be
\ba
\Pi_{B_1}(z_1, z_2;q)&=-{1\over 8}\left( \log^2 z_1  -2 \log z_1\log z_2 -\log^2 z_2 \right) +{1\over 2} \log z_2\, \widetilde \Pi_A (z_1, z_2;q)\\
&+ {1\over 4} \widetilde \Pi_B (z_1, z_2;q).
\ea
\ee
As noticed in \cite{acdkv}, the quantum period $\widetilde{\Pi}_{B}(z_1, z_2;q)$ can be computed by first extracting the finite part of the integral
\be
-16 \int_\delta^\Lambda \frac{\rd X}{X} \log V(X),
\ee
where $\delta$, $\Lambda$ are cut-offs and $V(X)$ is defined in (\ref{psi-psi}), and then by symmetrizing w.r.t. the exchange $\hbar \leftrightarrow -
\hbar$. By using the explicit solution for $V(X)$, we find
\be
\ba
\widetilde{\Pi}_{B}(z_1, z_2;q)&=8 \left[\frac{q+1}{2(q-1)} \log q \right] z_1+4 \left[ 1+\frac{5 q^2 + 8 q + 5}{2(q^2-1)}\log q \right] z_1^2 
\\ 
& +8 \left[ 1+\frac{(1+q)^3}{2 q(q-1)}\log q \right]z_1z_2
+4 z_2^2+{\cal O}(z_i^3).
\ea
\ee
In the classical limit $q\rightarrow 1$, one recovers the classical period defined in (\ref{Bper}):
\be
\lim_{q \to 1} \widetilde{\Pi}_{B}(z_1, z_2;q)=\widetilde{\Pi}_{B}(z_1, z_2).
\ee

Let us now consider the specialization (\ref{eq:z1z2-z}), and let us symmetrize w.r.t. $z_1$, $z_2$. We obtain
\be
\ba
&\widetilde{\Pi}_{B}\left(q^{\frac{1}{2}}z,q^{-\frac{1}{2}}z;q \right)
+\widetilde{\Pi}_{B} \left(q^{-\frac{1}{2}}z,q^{\frac{1}{2}}z;q\right)\\
&= \frac{4 (q+1)^2 \log (q)}{(q-1) \sqrt{q}} z  +
\Biggl[ \frac{2 \left(3 q^2+4 q+3\right)^2 \log (q)}{q \left(q^2-1\right)}+\frac{8 (q+1)^2}{q}\Biggr] z^2 \\
&+\Biggl[\frac{4 \left(9 q^8+79 q^7+222 q^6+405 q^5+482 q^4+405 q^3+222 q^2+79 q+9\right) \log (q)}{3 q^{5/2}
   \left(q^3-1\right)} \\
   & \qquad + \frac{8 \left(5 q^3+13 q^2+13 q+5\right)}{q^{3/2}}\Biggr] z^3+ \CO(z^4). 
\ea
\ee
As in the A-period case, and after using the identifications (\ref{eq:q-k}) and (\ref{z-mu}), we find that, up to an overall factor $1/(4 \pi^2 k)$, 
the coefficients in this expansion are precisely the coefficients $b_\ell(k)$ appearing in (\ref{b-bare}). We then propose the following identification:
\be
\label{jb-B}
J_b(\mu,k)=\frac{1}{4 \pi^2 k}\left(\widetilde{\Pi}_{B}\left(q^{\frac{1}{2}}z,q^{-\frac{1}{2}}z;q \right)
+\widetilde{\Pi}_{B} \left(q^{-\frac{1}{2}}z,q^{\frac{1}{2}}z;q\right)\right).
\ee
We have verified this equality up to sixth order in $z$. 

We then conclude that the problem of computing the membrane instanton corrections 
to the partition function of ABJM theory is completely solved by the above conjectural equivalence with 
quantum periods, i.e. with the refined topological string in the NS limit. 
Note that the natural solution of the Schr\"odinger equation that we have presented here, following \cite{acdkv}, is an expansion in $z$ but {\it exact} in $\hbar$. 
This is precisely what is needed for the M-theory expansion of the ABJM partition function, since it corresponds to an expansion at large $N$ but 
exact in the geometric parameter $k$. In particular, we can systematically compute the membrane instanton
corrections $a_\ell(k)$ and $b_\ell(k)$ by using the connection with
the refined topological string. 

It is important to notice that the worldsheet instanton expansion involves the quantum parameter 
\be
q_s =\re^{ 4 \pi \ri \over k}, 
\label{qWS}
\ee
while the membrane instanton expansion involves the quantum parameter
\be
q= \re^{\pi \ri k}. 
\label{qM2}
\ee
Therefore, we have some sort of $S$-duality acting on the coupling $1/k$. We will comment on this issue in section 4. 

As reviewed in subsection \ref{ref-rev}, the quantum periods can be used to determine the topological string free energy in the NS limit. We will now see how 
this point of view explains some of the structures discovered in \cite{hmo2}, like the appearance of an ``effective" chemical potential and the function $\widetilde J_b (\mu, k)$. 
By using (\ref{q-pers}) as well as (\ref{B1per}), one finds that the full free energy in the NS limit, $F_{\rm NS} (Q_1,Q_2;q)$, is defined by the equations, 
\be
\ba
Q_1 \partial_{Q_1} F_{\rm NS} (Q_1,Q_2;q)
&=-{1\over 4 \hbar} \left(\widetilde{\Pi}_{B}(z_1,z_2;q)
-\widetilde{\Pi}_{A}(z_1,z_2;q)^2\right),\\
Q_2 \partial_{Q_2} F_{\rm NS} (Q_1,Q_2;q)
&=-{1\over 4 \hbar}\left(\widetilde{\Pi}_{B}(z_2,z_1;q)
-\widetilde{\Pi}_{A}(z_2,z_1;q)^2\right). 
\label{dfns}
\ea
\ee
Here we only consider the instanton part of the free energy, and we dropped quadratic terms in the moduli $T_1$, $T_2$ which are not relevant for our purposes. By using the explicit 
results for the quantum periods listed above, one finds
\be
F_{\rm NS}(Q_1, Q_2;q) = -{1+ q \over q-1}  \left(Q_1 + Q_2 \right)-\frac{q^2+1}{4 \left(q^2-1\right)} \left( Q_1^2 + Q_2^2\right) -\frac{(q+1)^2 \left(q^2+1\right)}{q \left(q^2-1\right)}Q_1 Q_2 + \cdots
\ee
One can extract from this expression the GV invariants (\ref{ns-invs}). They agree, up to an overall sign $(-1)^g$, with the results listed in \cite{ckk}. 

We are now ready to interpret the relationship (\ref{mueff-mu}) in the light of the refined topological string. 
As we have shown, the coefficients $a_\ell(k)$ are, up to an overall constant, the 
coefficients of the quantum mirror map, evaluated at the slice (\ref{eq:z1z2-z}). It is easy to see that this slice in the $z_1-z_2$ space corresponds, in terms of the $Q_I$ 
variables, $I=1,2$, to a slice 
which we can parametrize as
\be
\label{q-slice}
Q_1=q^{1/2}Q, \qquad Q_2=q^{-1/2}Q.
\ee
Since $\mu$ corresponds to the variable $z$ in the moduli space, through the identification $z=\re^{-2\mu}$, 
we find from (\ref{mueff-mu}) that the flat coordinate $Q$ is given by,  
\be
\label{Qmueff}
Q=\re^{-2\mu_{\rm eff}},
\ee
Therefore, the relationship between the effective chemical potential (\ref{mueff-mu}), which incorporates bound states in the ABJM partition function, and the ``bare" chemical potential $\mu$ 
is just the quantum mirror map. 

Using now (\ref{A-ja}) and (\ref{jb-B}), as well as (\ref{dfns}), we find that the combination defining the function $\widetilde{J}_b(\mu_{\rm eff},k)$, and introduced in the first line of (\ref{widetildeJ}), is precisely 
what is needed to obtain the symmetric combination of quantum periods, 
\be
\label{Jbrefined}
\widetilde{J}_b(\mu_{\rm eff},k)
=-\frac{\ri}{\pi}
\left(Q_1 \partial_{Q_1}F_{\rm NS} \left( Q_1,Q_2;q \right)+Q_2 \partial_{Q_2}F_{\rm NS}\left(Q_1,Q_2;q\right)\right),
\ee
evaluated at the slice (\ref{q-slice}), and where the variables in the l.h.s. are related to those in the r.h.s by (\ref{Qmueff}) and (\ref{q-k}).

As we reviewed in (\ref{ref-rev}), the refined topological string free energy can be expressed in terms of refined GV invariants, which in the case of 
local $\IP^1 \times \IP^1$ depend on two degrees, $n_{g_L, g_R}^{d_1, d2}$.  They have the symmetry property
\be
n_{g_L, g_R}^{d_1, d_2}=n_{g_L, g_R}^{d_2, d_1}.
\ee
In the NS limit, the free energy of local $\IP^1 \times \IP^1$ has the integrality structure (\ref{NS-GV}). This, together with (\ref{Jbrefined}), 
gives the following expression for the coefficients $\widetilde b_\ell(k)$ of $\widetilde{J}_b(\mu_{\rm eff},k)$ in terms of refined GV
invariants:
\be
\label{bNS}
\widetilde b_\ell(k)= -{\ri \over \pi} \sum_{g\ge 0} \sum_{d|\ell} 
{d^2 \over \ell} \hat n_g^d ( q^{\ell/d}) 
{\left(q^{\ell /4 d}-q^{-\ell /4 d}\right)^{2g} \over q^{\ell/2d}-q^{-\ell/2d}}. 
\ee
Here, 
\be
\hat n_g^d ( q) = \sum_{d_1+d_2=d} \hat n_g^{d_1, d_2} q^{(d_1-d_2)/2},
\ee
and the invariants $\hat n_g^{d_1, d_2} $ are defined in (\ref{ns-invs}). We can use explicit results for the refined GV invariants to check that the 
above expression matches with \eqref{b1b2b3}.

\subsection{The HMO cancellation mechanism in terms of BPS invariants}

As we have reviewed in section 2, the partition function of ABJM theory, i.e. the matrix 
integral (\ref{abjmmatrix}), is finite for any value of $k$. This means that the poles appearing at integer values of $k$ in 
the worldsheet instanton contribution $J^{\rm WS}(\mu, k)$ should be cancelled by poles appearing in the membrane instanton and 
bound state contribution. This is the HMO cancellation mechanism discovered in \cite{hmo1}. 

As noticed in \cite{hmo2}, in order to study the cancellation mechanism 
in full generality, it is convenient to look at the expression (\ref{gpmueff}). Since the perturbative part is regular for any $k$, we have to make sure that 
the poles appearing in $J^{\rm WS}(\mu_{\rm eff}, k)$ are cancelled by similar poles in the last two terms of (\ref{gpmueff}). 
In previous subsections we have shown that the contribution of membrane instantons can be written in terms of the free energy of the refined topological string. In particular, 
in \eqref{Jbrefined} and (\ref{bNS}) we have related the function $\widetilde J_b (\mu_{\rm eff},k)$ to refined GV invariants. By the conjectured equality (\ref{bcrel}), 
the function $\widetilde J_c (\mu_{\rm eff},k)$ can 
be also written in terms of these invariants. In this subsection we will show that the conjectures (\ref{bNS}) and (\ref{bcrel}) make it possible to explain the 
cancellation mechanism by using the refined integrality structure as well as properties of the refined BPS invariants. 

In order to proceed, it is more elegant to express the relevant functions in terms of the BPS indices $N_{j_L, j_R}^{d_1, d_2}$ 
of the local $\IP^1 \times \IP^1$ geometry (a list of these invariants for the first few degrees $d_1$, $d_2$ can be found in \cite{ckk}). 
Using the GV integrality (\ref{gv-exp}), together with (\ref{change-basis}), the worldsheet instanton part 
can be written as 
\be
\label{ws-ds}
 J^{\rm WS}(\mu_{\rm eff} ,k)=\sum_{m \ge 1}d_m(k) \re^{-4 m \mu_{\rm eff}/k },
 \ee
 where 
 \be
 \label{dmk}
 d_m(k)=\sum_{j_L,j_R}\sum_{m=dn}\sum_{d_1+d_2=d}N^{d_1,d_2}_{j_L,j_R}
\frac{2j_R+1}{(2\sin\frac{2\pi n}{k})^2}\frac{\sin\left( \frac{4\pi n}{k}(2j_L+1) \right)}{\sin\frac{4\pi n}{k}}\frac{(-1)^m}{n}. 
\ee
On the other hand, we can rewrite (\ref{bNS}) as 
\be
\label{blj}
\widetilde{b}_\ell(k)=-\frac{\ell}{2\pi}\sum_{j_L,j_R}\sum_{\ell=dw}\sum_{d_1+d_2=d}N^{d_1,d_2}_{j_L,j_R}q^{\frac{w}{2}(d_1-d_2)}
\frac{\sin\frac{\pi kw}{2}(2j_L+1)\sin\frac{\pi kw}{2}(2j_R+1)}{w^2\sin^3\frac{\pi kw}{2}}. 
\ee
Here, $n$ and $w$ denote the multi-covering numbers for the worldsheet instanton
and membrane instanton, respectively. 

The HMO cancellation mechanism states that, order by order in $\re^{-\mu_{\rm eff}}$, 
the total grand potential (\ref{gpmueff}) should be regular. The coefficient \eqref{dmk} has double poles when $k \in 2n/\IN$. The coefficient (\ref{blj}) has 
a simple pole when $k \in 2 \IN/w$, and due to (\ref{bcrel}) 
the coefficient $\tilde c_\ell (k)$ will have a double pole at the same values of $k$. These poles contribute to terms of the same order in $\re^{-\mu_{\rm eff}}$
precisely when $k$ takes the form
\be
\label{ksing}
k={2n \over w}={2m \over \ell}.
\ee
 We have then to examine the pole structure of (\ref{gpmueff}) at these values of $k$. Since both (\ref{dmk}) and (\ref{blj}) involve a sum over BPS multiplets with quantum numbers given 
 by degrees $(d_1, d_2)$ and spins $(j_L, j_R)$, we can look at the contribution to the pole structure of each multiplet. In the worldsheet instanton contribution, the singular part associated to a BPS multiplet around $k=2n/w$ is 
 given by 
 \be
 \label{ws-pole}
{(-1)^m \over \pi^2}  \left[  {n\over  w^4 \left( k -{2n \over w}\right)^2 } + {1\over    k -{2n \over w}}  \left( {1\over w^3} +   {m \over  n w^2}   \mu_{\rm eff}   \right) \right] (1+2 j_L)(1+ 2j_R) 
N^{d_1,d_2}_{j_L,j_R} \re^{-{2 mw\over n}  \mu_{\rm eff} }.
 \ee
The singular part in $\mu_{\rm eff} \widetilde J_b (\mu_{\rm eff},k) $ associated to a BPS multiplet is given by 
\be
\label{b-pole}
-{  \re^{\pi \ri k w(d_1-d_2)/2} \over \pi^2}  {\ell \over w^3 \left( k -{2n \over w} \right)} (-1)^{n(2j_L+2j_R-1)} (1+2 j_L)(1+ 2j_R)  N^{d_1,d_2}_{j_L,j_R}\mu_{\rm eff}\re^{-2\ell\mu_{\rm eff}}. 
\ee
Using (\ref{bcrel}), we find that the corresponding singular part in $\widetilde J_c (\mu_{\rm eff},k) $ is given by 
\be
\label{c-pole}
-{\re^{\pi \ri k w(d_1-d_2)/2} \over \pi^2} \left[  {n\over  w^4 \left( k -{2n \over w}\right)^2 } + {1\over    w^3 \left( k -{2n \over w}  \right)} \right] (-1)^{n(2j_L+2j_R-1)} (1+2 j_L)(1+ 2j_R)  N^{d_1,d_2}_{j_L,j_R}\re^{-2\ell\mu_{\rm eff}}
\ee
By using (\ref{ksing}), one notices that 
\be
 \re^{\pi \ri k w(d_1-d_2)/2}=(-1)^m
\ee
and it is easy to see that all poles in (\ref{ws-pole}) cancel against the poles in (\ref{b-pole}) and (\ref{c-pole}), for any value of $\mu_{\rm eff}$, provided that
\be
(-1)^{n(2j_L+2j_R-1)}=1.
\ee
However, this is the case, since for local $\IP^1 \times \IP^1$ the only non-vanishing BPS indices $N^{d_1, d_2}_{j_L, j_R}$ have 
\be
\label{2jzero}
2j_L + 2j_R-1\equiv 0 \quad {\rm mod}\, 2. 
\ee
This can be justified by the following geometric argument.\footnote{We are thankful to Albrecht Klemm for explaining this argument to us.} The spins $j_L$ and $j_R$ are 
related to the Lefshetz decomposition of the cohomology of the moduli space of an M2-brane wrapping the two-cycle with degree ${\bf d}$ \cite{gv,kkv}. We will denote 
the cohomology class of this cycle by $C$. This moduli space 
is, as described in \cite{gv,kkv}, a torus fibration $\IT^{2g}$ over the geometric deformation space $\CM_C$ of the two-cycle, where $g$ is 
the genus of a smooth curve in the class $C$. The maximal value of $2j_L$ is given by $g$, while the maximal value of $2j_R$ is given by the dimension of the moduli space $\CM_C$. 
This space can be taken to be $|C| \cong \IP H^0(\CO(C))$, the complete linear system associated to $C$. Using the adjunction formula together with Riemann--Roch, we find \cite{kkv}
\be
\ba
 2j_L^{\rm max}&=g(C)=\frac{C^2+KC}{2}+1, \\
2j_R^{\rm max}&=h^0(\mathcal{O}(C))-1=g(C)+d(C)-1, 
\ea
\ee
where
\be
d(C)=-KC
\ee
 is the degree of the curve $C$ w.r.t. the anti-canonical class. We then find, 
\be 
2j_L^{\rm max}+2j_R^{\rm max}-1=2g(C)-2+d(C)
\ee
and
\be
(-1)^{2j_L+2j_R-1}=(-1)^{d(C)}.
\label{jtod}
\ee
Now, in the case of local $\IP^1 \times \IP^1$, a curve in the class $(d_1, d_2)$ has degree w.r.t. to the 
anticanonical class given by $d(C)=2 d_1 + 2 d_2$, which is even. (\ref{2jzero}) follows, and this guarantees the cancellation 
of the poles. 

We conclude that the relation between membrane instantons and quantum periods that we have conjectured guarantees the 
HMO cancellation mechanism and implements it through the properties of the refined BPS invariants. 

\subsection{Analytic properties of the grand potential}

Now that we have completely determined the structure of the grand potential, including all non-perturbative corrections, it is interesting to ask what are its properties as an expansion 
at large $\mu$. Since we are interested in the M-theory point of view, 
we want to understand the behavior of the power series expansions in $\re^{-2\mu}$, $\re^{-4 \mu/k}$ (i.e. at large $N$) for $k$ fixed. In particular, we want to 
determine how the coefficients $a_\ell (q)$, $b_\ell(q)$, $c_\ell(q)$ in (\ref{gen-M2}) and $d_\ell(q_s)$ in (\ref{ws-ds}) grow with $\ell$ . We have 
written their dependence w.r.t. $k$ in terms of the variables $q$ and $q_s$, defined in (\ref{qM2}) and (\ref{qWS}), respectively. 

\FIGURE{
\includegraphics[height=3.5cm]{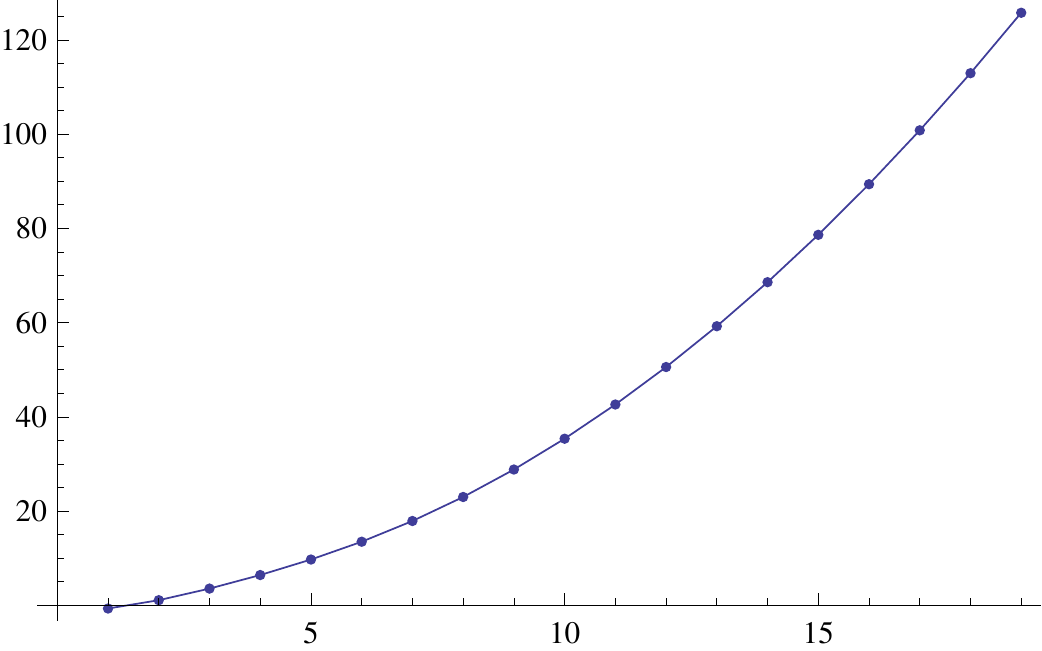} \qquad \qquad  \includegraphics[height=3.5cm]{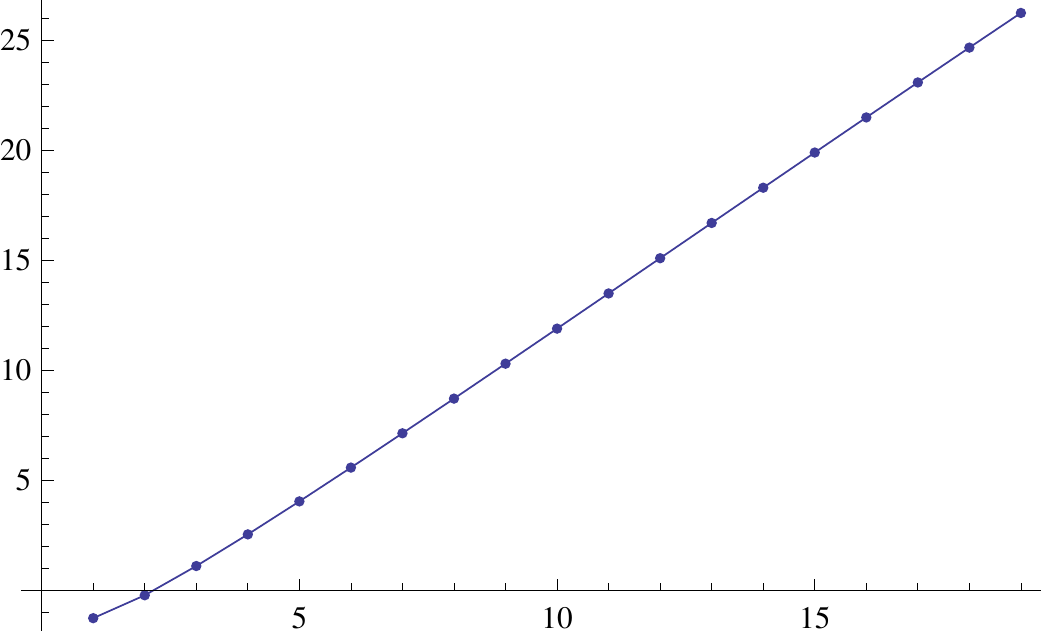} 
\caption{A plot of $\log |a_\ell(q)|$ for $\ell=1, \cdots, 20$, evaluated at $q=1/4$ (left) and $q=\exp( \pi \ri /2)$ (right), displaying the behaviors (\ref{divC}) and (\ref{ling}), respectively.}
\label{graphics-a}
}
Using the values of these coefficients, we have found an interesting pattern. It turns out that their growth with $\ell$ depends crucially on the 
value of $q$ and $q_s$, regarded as complex parameters. When $|q|, |q_s| \not=1$, these coefficients grow like 
\be
\label{divC}
\sim \exp (C  \ell^2), \qquad \ell \gg 1. 
\ee
However, when $|q|, |q_s| =1$ we have the milder growth 
\be
\label{ling}
\sim \exp (C  \ell), \qquad \ell \gg 1. 
\ee
As an example, we show in \figref{graphics-a} 
the growth of $\log |a_\ell(q)|$ for $\ell=1, \cdots, 20$ for two values of $q$ (one with $|q|\not=1$ and another one with $|q|=1$), 
which illustrate our claim. When $k=2n$ is an even integer, one can 
compute the generating functional of the $a_\ell (k)$ explicitly \cite{hmo2}, and one finds 
\be
\sum_{\ell=1}^\infty  a_\ell (2n) \re^{-2 \ell \mu}= {2 (-1)^{n-1} \over n \pi^2} \re^{-2 \mu} {}_4F_3 \left( 1,1, {3\over 2}, {3\over 2}; 2,2,2; (-1)^n 16 \re^{-2 \mu} \right),
\ee
which confirms the growth (\ref{ling}). Notice that, when $q$ and $q_s$ are roots of unity (in particular, when they are of the form (\ref{qM2}) and (\ref{qWS}) and $k$ is an integer), the coefficients $b_\ell(q)$, $c_\ell(q)$ and $d_\ell(q_s)$ have poles for an infinite subsequence of values of $\ell$. We find however that the growth of the coefficients which are finite is still of the form (\ref{ling}). 
\FIGURE{
\includegraphics[height=3.5cm]{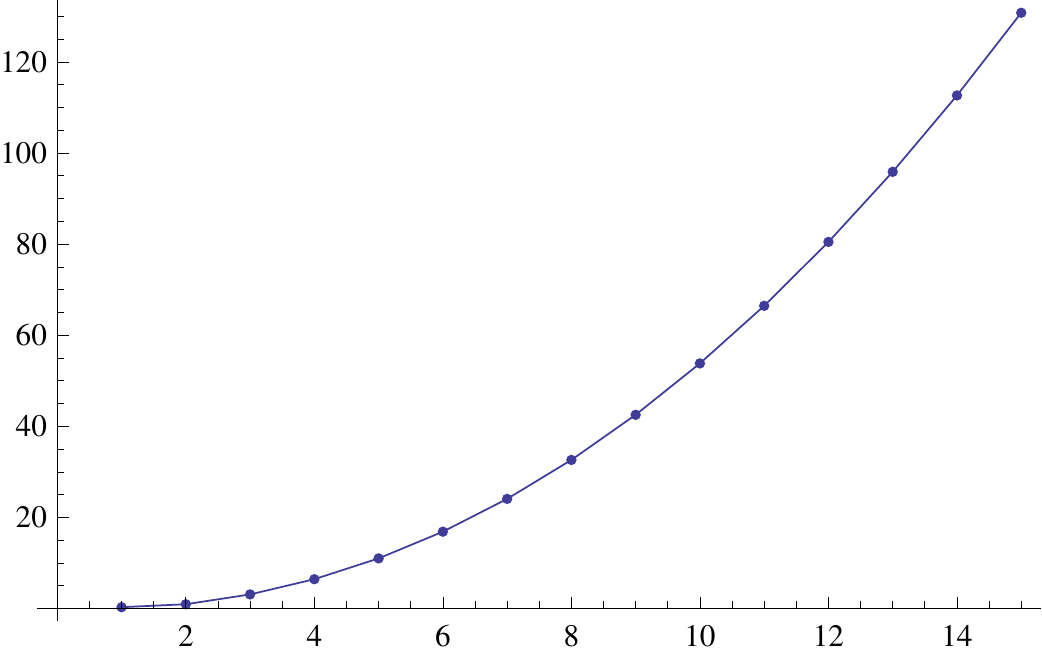} \qquad \qquad \includegraphics[height=3.5cm]{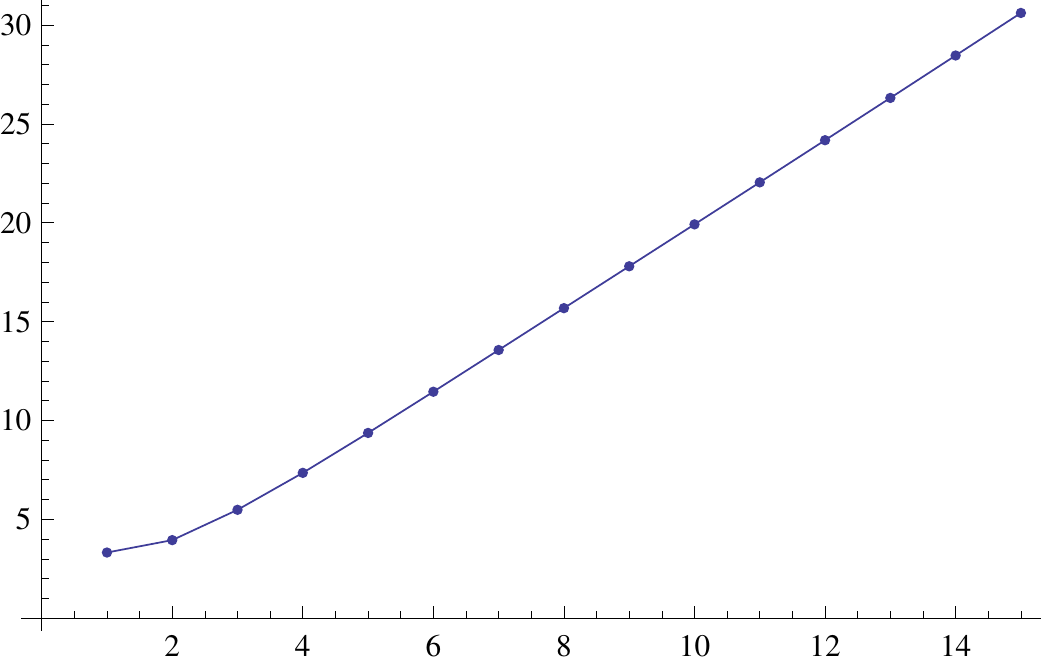} 
\caption{A plot of $\log |d_\ell(q_s)|$ for the topological string on local $\IP^2$ and $\ell=1, \cdots, 15$, and evaluated at $q_s=1/4$ (left) and $q_s=\exp(2 \pi \ri /19)$ (right). They display the behaviors (\ref{divC}) and (\ref{ling}), respectively.}
\label{graphics-d}
}

We have verified that our observation is also valid for the coefficients $d_\ell (q_s)$ appearing in the Gopakumar--Vafa expansion of the topological 
string free energy in other models. In \figref{graphics-d} we show the growth of $\log |d_\ell(q_s)|$ for $\ell=1, \cdots, 15$ in local $\IP^2$, for two values of $q$, 
and we conjecture that this is a general feature of local CYs. It is also natural to conjecture that the growth properties of the coefficients $a_\ell(q)$, $b_\ell(q)$ that we have found in local $\IP^1 \times \IP^1$ will be also found in the coefficients of the quantum periods of any local CYs. 

The main consequence of this conjectural growth is that the expansion (\ref{jnpgen}) is not just an asymptotic expansion for $\mu \gg 1$: when $k$ is an integer, as required in ABJM theory, 
the expansion has actually a {\it finite} radius of convergence around $\mu =\infty$. This can be seen for example by looking at the expressions (\ref{mueff-mu}) and (\ref{conj-one}): if our conjecture is true, (\ref{mueff-mu}) defines $\mu_{\rm eff}$ as an analytic function of $\re^{-\mu}$, around $\mu=\infty$, while (\ref{conj-one}) defines an analytic function of $\mu_{\rm eff}$

It was argued in \cite{pv} that the coefficients of membrane instanton generating functionals should grow like (\ref{divC}). Here we find, however, that this generic behavior becomes milder for integer (and even real) $k$. The growth (\ref{divC}) was used in \cite{pv} to argue for the existence of non-perturbative corrections due to 5-branes, by using the standard argument on 
the ambiguities associated to asymptotic series. Since in our case the grand potential is given by a convergent series, this non-perturbative ambiguity is absent.

\section{A proposal for non-perturbative topological strings}

In the previous section, we have seen that the coefficients $a_\ell(k)$
and $b_\ell(k)$ (or $\widetilde b_\ell(k)$)
of membrane instantons are given by the
quantum A- and B-periods, respectively.
Therefore, if we also use (\ref{bcrel}), we see that membrane instanton corrections are
determined by the refined topological strings in the NS limit.
We have also seen that poles coming from worldsheet instantons and
membrane instantons cancel with each other.
This cancellation does not depend on the
details of the invariants $N^{\bf d}_{j_L,j_R}$,
and the cancellation occurs within each BPS multiplet.

This suggests that we can generalize the pole cancellation mechanism
for arbitrary local CY,
and find the non-perturbative completion of topological string partition functions
using the HMO mechanism as a guiding principle.
In fact, it is natural to expect that the partition function is a smooth function of 
the string coupling
so that we can go smoothly
from the weak coupling to the strong coupling: this is basically the raison d'\^{e}tre of M-theory. 
The analysis in \cite{hmo1,cm,hmo2} shows that the pole
cancellation gives a very strong constraint which almost determines
the expression of free energy.

\subsection{From ABJM theory to arbitrary local Calabi--Yau manifolds}
In order to generalize
the expression of the ABJM grand potential
to arbitrary local CY, let us take a closer look at 
the membrane instanton part of the ABJM grand potential \eqref{gpmueff}
\begin{align}
 J^{\rm M2}(\mu_{\rm eff},k)=\sum_{\ell=1}^\infty
\left[\widetilde{b}_\ell(k)\mu_{\rm eff}-k^2\frac{\partial}{\partial k}
\left(\frac{\widetilde{b}_\ell(k)}{2\ell k}\right)\right]\re^{-2\ell\mu_{\rm eff}},
\label{JM2b}
\end{align}
where we have used the relation between $\widetilde{c}_\ell(k)$ and 
$\widetilde{b}_\ell(k)$ in \eqref{bcrel}.
Notice that
$-k^2\partial_k$ is essentially the derivative w.r.t. the string coupling constant $g_s$ in (\ref{relatio}).
As we will see below, it is convenient to 
introduce a  different 
normalization of the string coupling constant
\begin{align}
 \lambda_s=\frac{2}{k}.
\end{align}
In terms of $\lambda_s$, the quantum parameters
for the worldsheet instantons \eqref{qWS}
and the membrane instantons \eqref{qM2}
are simply related by the
inversion of the coupling
$\lambda_s\rightarrow1/\lambda_s$
\begin{align}
 q_s=\re^{\frac{4\pi \ri}{k}}=\re^{2\pi \ri\lambda_s},\quad
q=\re^{\ri\pi k}=\re^{\frac{2\pi \ri}{\lambda_s}}.
\label{qsq}
\end{align}
Also, one can easily see that $J^{\rm M2}$ can be written
as a total derivative w.r.t. $\lambda_s$ if we treat the parameter
\be
\CT=\frac{4\mu_{\rm eff}}{k}
\ee
and $\lambda_s$ as independent variables, i.e.
$\partial \CT/\partial\lambda_s=0$. We find, 
\begin{align}
 J^{\rm M2}
=\frac{\partial}{\partial \lambda_s}\left[\lambda_s\sum_{\ell=1}^\infty
\frac{\widetilde{b}_\ell(k)}{2\ell }\re^{-\ell\frac{\CT}{\lambda_s}}\right].
\label{JM2del}
\end{align}

We can further rewrite $J^{\rm M2}$ 
in a form which is more suitable for a generalization to arbitrary local CY.
Plugging the expansion of 
$\widetilde{b}_\ell(k)$ \eqref{blj}
into \eqref{JM2del}, one finds that 
$J^{\rm M2}$ is written as a derivative of 
the refined free energy in the NS limit\footnote{
In this section, we use a slightly different notation
for the free energies of the ordinary (unrefined) topological string
\eqref{gv-exp} and
the refined topological string in the NS limit \eqref{NS-j}
\begin{align*}
 F_{\rm top}({\bf T},\lambda_s)&=-\sum_{n=1}^\infty\sum_{{\bf d}}\sum_{j_L,j_R}
N_{j_L,j_R}^{{\bf d}}
\frac{(2j_R+1)\chi_{j_L}(q_s^{n})}{\left(q_s^{n/2}-q_s^{-n/2}\right)^2}
\frac{\re^{-n{\bf d}\cdot {\bf T}}}{n},\\
F_{\rm NS}\left(\frac{{\bf T}}{\lambda_s},\frac{1}{\lambda_s}\right)&=
\sum_{w=1}^\infty\sum_{{\bf d}}\sum_{j_L,j_R}
N_{j_L,j_R}^{{\bf d}}
\frac{\chi_{j_L}(q^{w/2})\chi_{j_R}(q^{w/2})}{q^{\frac{w}{2}}-q^{-\frac{w}{2}}}
\frac{\re^{-w\frac{{\bf d}\cdot {\bf T}}{\lambda_s}}}{w^2},
\end{align*}
with $q_s$ and $q$ given by \eqref{qsq}.
}
\begin{align}
 J^{\rm M2}=\frac{1}{2\pi\ri}\frac{\partial}{\partial \lambda_s}\left[\lambda_s F_{\rm NS}\left(\frac{T^{\rm eff}_1}{\lambda_s},\frac{T^{\rm eff}_2}{\lambda_s},\frac{1}{\lambda_s}\right)\right].
\label{JM2NS}
\end{align}
Here $T^{\rm eff}_1$ and $T^{\rm eff}_2$ denote
\be
\label{T1T2}
T^{\rm eff}_1=\frac{4\mu_{\rm eff}}{k}-\ri\pi, \qquad T^{\rm eff}_2=\frac{4\mu_{\rm eff}}{k}+\ri\pi. 
\ee
 The derivative w.r.t. $\lambda_s$ in (\ref{JM2NS}) is again taken by assuming that $T_i^{\rm eff}$ are independent of $\lambda_s$. 
The parameters $T_i^{\rm eff}$, $i=1,2$, can be also written in terms of the K\"ahler parameter $T$ of local $\IP^1 \times \IP^1$ in the diagonal slice, which was defined in (\ref{relatio}), as 
\be
\label{T1T2T}
\ba
 T^{\rm eff}_1&= T -\lambda_s \widetilde \Pi_A \left( \frac{T}{\lambda_s},\frac{T + 2 \pi \ri }{\lambda_s};q\right) ,\\
 T^{\rm eff}_2&= T + 2 \pi \ri -\lambda_s \widetilde \Pi_A \left( \frac{T}{\lambda_s},\frac{T+ 2 \pi \ri }{\lambda_s};q\right).
 \ea
 \ee
 The notation in the arguments of $ \widetilde \Pi_A$ in (\ref{T1T2T}) means that the quantum A-period (\ref{Aper}) is evaluated at 
 \be
 \label{zsnp}
 z_1=\exp\left(-{T\over \lambda_s}\right), \qquad z_2=\exp\left(-{T+ 2 \pi \ri \over \lambda_s}\right). 
 \ee
 There are two important remarks to be made on the effective K\"ahler parameters introduced in (\ref{T1T2T}). First of all, 
 they differ from the conventional K\"ahler parameters by non-perturbative terms, given by the quantum A-period, which are needed to take into 
 account bound states. This period 
 is not evaluated on the usual complex deformation parameters, but on the variables (\ref{zsnp}), 
 which are non-perturbative and not analytic as $\lambda_s=0$. 
 Second, since K\"ahler parameters are defined only modulo $2 \pi \ri$ (this is the periodicity of the $B$-field), 
 the relative shift between $T_1^{\rm eff}$ and $T_2^{\rm eff}$ in (\ref{T1T2}) 
is not visible in the worldsheet instanton sector, and the ``effective" K\"ahler parameters $T_i^{\rm eff}$, $i=1,2$, 
still belong to the diagonal slice of local $\mathbb{P}^1\times \mathbb{P}^1$ in the sense that 
\begin{align}
 \re^{-T^{\rm eff}_1}=\re^{-T^{\rm eff}_2}.
\end{align}
Thus, the worldsheet instanton part of
the ABJM grand potential \eqref{gpmueff} is also
written in terms of the
$T^{\rm eff}_{1},T^{\rm eff}_2$ in \eqref{T1T2}, as
\begin{align}
 J^{\rm WS}(\mu_{\rm eff},k)=F_{\rm top}(T^{\rm eff}_1,T^{\rm eff}_2,\lambda_s),
\end{align}
where $F_{\rm top}$ is the (unrefined) topological string free energy. 
Finally, we arrive at an elegant formula for the
non-perturbative part of the
ABJM grand potential
\begin{align}
 J^{\rm (np)}(\mu_{\rm eff},k)=F_{\rm top}(T^{\rm eff}_1,T^{\rm eff}_2,\lambda_s)
+\frac{1}{2\pi\ri}\frac{\partial}{\partial \lambda_s}\left[\lambda_s F_{\rm NS}\left(\frac{T^{\rm eff}_1}{\lambda_s},\frac{T^{\rm eff}_2}{\lambda_s},\frac{1}{\lambda_s}\right)\right].
\label{JnpFref}
\end{align}
Note as well that the shift in $T^{\rm eff}_2$ w.r.t. $T^{\rm eff}_1$ by $2 \pi \ri$ units is relevant in the membrane instanton sector, since 
there this shift is divided by $\lambda_s$. Therefore, it also 
appears in the arguments of the quantum period $\widetilde \Pi_A$ in (\ref{T1T2}). 

As we mentioned in the introduction, the results of \cite{mp,dmp} indicate that the ABJM matrix model derived in \cite{kwy} 
provides a non-perturbative definition of topological string theory on local $\IP^1 \times \IP^1$. Here we are restricting ourselves to the ABJM slice 
where the gauge groups of ABJM theory have the same rank, $N_1=N_2=N$, and so we obtain local $\IP^1\times \IP^1$ along the diagonal direction. 
The grand potential of ABJM theory can then be interpreted as the {\it non-perturbative} topological string free energy of local $\IP^1 \times \IP^1$, in the large radius 
frame. Our result in (\ref{JnpFref}) can be then interpreted as a calculation of the expansion of the non-perturbative 
free energy, along the slice $T_1=T_2=T$, for ${\rm Re}(T)\gg 1$, and $\lambda_s>0$ (in particular, the topological string coupling is imaginary). Therefore, we have
\be
F^{({\rm np})}(T, \lambda_s)=  F_{\rm top}(T^{\rm eff}_1,T^{\rm eff}_2,\lambda_s)
+\frac{1}{2\pi\ri}\frac{\partial}{\partial \lambda_s}\left[\lambda_s F_{\rm NS}\left(\frac{T^{\rm eff}_1}{\lambda_s},\frac{T^{\rm eff}_2}{\lambda_s},\frac{1}{\lambda_s}\right)\right].
\label{FnpFref}
\ee
where the relationship between $T^{\rm eff}_i$, $i=1,2$ and $T$ is spelled out in detail in (\ref{T1T2T}). 

We want to emphasize that this is a {\it first-principles calculation} of the expansion of the free energy at large $T$, including the full series of non-perturbative corrections. 
Although our formula (\ref{FnpFref}) is conjectural, it agrees with a large amount of data 
concerning the matrix model, as we have explained in the previous section of this paper. Let us make some remarks on the structure of the answer (\ref{FnpFref}). 

\begin{enumerate}

\item It contains terms which are not analytic at $\lambda_s=0$, of the form 
\be
\re^{-T/\lambda_s}. 
\ee
These non-perturbative terms are encoded in the quantum periods, therefore they are determined by the NS limit of the topological string. They 
appear both as corrections to the K\"ahler parameters in (\ref{T1T2T}), and in the NS free energy in (\ref{FnpFref}). They seem to correspond to some sort of 
``topological membrane instantons." Also, the coefficients appearing in the quantum 
periods are functions of $q=\re^{2\pi \ri /\lambda_s}$ 
are therefore are not analytic at $\lambda_s=0$.

\item In the limit $\lambda_s \rightarrow 0$, the non-perturbative corrections drop out and we are left with the perturbative, 
topological string free energy $F_{\rm top}(T,T,\lambda_s)$ along the diagonal of local $\IP^1 \times \IP^1$. 

\item The quantum A-periods lead to mixed terms of the form, 
\be
\re^{-n T/\lambda_s - m T}
\ee
which seem to correspond to bound states of perturbative worldsheet instantons and membrane instantons, as in the M-theory dual to ABJM theory. 

\item The total free energy satisfies the HMO cancellation mechanism, i.e. there are no poles in $\lambda_s$, order by order in $\re^{-T}$. 
\end{enumerate}

Notice that, for general, complex values of $\lambda_s$, the expansion we have found is an asymptotic expansion at large $T$, and the form of the expansion might 
change along different directions of the complex plane due to the Stokes phenomenon. This was 
displayed in detail in a closely related situation in \cite{mpp}. 

Since the expression of $F^{({\rm np})}(T, \lambda_s)$ is 
completely written in the language of refined topological strings, 
it seems straightforward to
generalize it to an arbitrary, local CY: the total free energy should be the sum of the standard, perturbative topological string free energy, evaluated at some ``effective" K\"ahler parameters, plus the 
derivative of the NS free energy appearing in (\ref{FnpFref}). However, for the pole cancellation mechanism to work,
we have to be careful about an extra minus sign
which was absent in the ABJM case.
By generalizing \eqref{JnpFref} to arbitrary local CY
and repeating a similar computation of the residue in
the previous section,
one can easily see that
the pole at $\lambda_s=w/n$ 
almost cancels between $F_{\rm top}$ and $F_{\rm NS}$,
except for a sign difference
$(-1)^{n(2j_L+2j_R-1)}$.
Here $n$ and $w$ denote the multi-covering numbers
in $F_{\rm top}$ and $F_{\rm NS}$, respectively.
This sign was absent in the local $\mathbb{P}^1\times \mathbb{P}^1$ case
since $(-1)^{2j_L+2j_R-1}=1$, while this is not the case in general
local CY. 
As shown in \eqref{jtod}, this sign is related to the
degree of the curve $d(C)=-KC$, where $K$ denotes the canonical class of the base of local CY.
Therefore, this extra sign can be taken care of 
by turning on a discrete B-field,
$B=\pi K$, along the worldsheet instantons
\begin{align}
\re^{-\ri\int_CB}=\re^{-\ri\pi KC}=(-1)^{2j_L+2j_R-1}.
\end{align}
Below, we denote this discrete B-field by ${\bf K}=(K_1,K_2,\dots)$, which plays a crucial role 
in the pole cancellation mechanism.
As above, the $I$-th component $K_I$ is given by the integral of $B$ 
along the two-cycle with complexified size $T_I$,
and takes a half-integer value.
The final answer will involve ``effective" K\"ahler parameters $T^{\rm eff}_I$, $I=1,\cdots, n$. 
These are allowed to differ from the perturbative K\"ahler parameters in a shift 
by another $B$-field $2 \pi \ri {\bf n}$, with ${\bf n}=(n_1, n_2,\dots)$, 
and a non-perturbative part involving the quantum A-periods. 
Here we propose the relationship of $T_I$ and $T_I^{\rm eff}$,
\be
\label{teffgen}
T_I ^{\rm eff}=T_I+ 2 \pi \ri n_I - \lambda_s \widetilde \Pi_{A_I} \left({T_I+ 2 \pi \ri n_I \over \lambda_s}; q \right), \qquad I=1, \cdots, n. 
\ee
In this equation and the following ones, the string coupling constant $\lambda_s$ is defined by (\ref{qsq}). The quantum A-periods are written as 
\be
\Pi_{A_I}(z_I;q)= \log z_I +  \widetilde \Pi_{A_I}(z_I;q), 
\ee
and the notation in (\ref{teffgen}) means that the $\widetilde \Pi_{A_I}(z_I;q)$ are evaluated at 
\be
z_I=\exp \left(-{T_I+ 2 \pi \ri n_I\over \lambda_s}  \right). 
\ee
In the diagonal, local $\IP^1 \times \IP^1$, one has the B-field shift $n_1=0$, $n_2=1$. 
Cancellation of the poles also require the $n_I$ to be integers, so that their presence is only visible in the membrane instanton sector. 
From this argument, we propose
the following expression of the non-perturbative
topological string free energy for an arbitrary local CY:
\begin{align}
 F^{\rm (np)}({\bf T}+ \pi \ri {\bf K},\lambda_s)=F_{\rm top}({\bf T}^{\rm eff}+ \pi \ri {\bf K},\lambda_s)+
\frac{1}{2\pi\ri}\frac{\partial}{\partial \lambda_s}\left[\lambda_s F_{\rm NS}\left(\frac{{\bf T}^{\rm eff}}{\lambda_s},\frac{1}{\lambda_s}\right)\right].
\label{ourFnp}
\end{align}
Notice that, when written in this form, the natural perturbative K\"ahler parameter (i.e. the K\"ahler parameter appearing in the worldsheet instanton part) is given by 
${\bf T}+ \pi \ri {\bf K}$ and includes the half-integer shift by ${\bf K}$. Equivalently, we can redefine the K\"ahler parameter in such a way that the shift appears in the membrane instanton 
part.

To summarize, as a natural generalization of
the ABJM grand potential \eqref{JnpFref}, we arrived at
our proposal \eqref{ourFnp} for the non-perturbative topological string
free energy
as a sum of the unrefined free energy and the refined free energy in the NS
limit. In addition, from the requirement of pole cancellation,
we have to turn on a discrete B-field flux $\pi \ri {\bf K}$ along the worldsheet instantons. There is in principle a non-trivial, ``non-perturbative" 
B-field $2 \pi \ri{\bf n}$ in the membrane instanton 
contribution which can not be fixed {\it a priori}, but this is the only unknown datum of our proposal.

\subsection{Toward non-perturbative refined topological strings}
Our result (\ref{FnpFref}), as well as our general proposal \eqref{ourFnp}, seem to be related to a recent 
suggestion by Lockhart and Vafa in \cite{lv}, where a proposal for the calculation of non-perturbative effects in refined topological strings
was put forward. This proposal is based on the formal similarity between the 
refined topological string partition function, 
and the triple sine function which appears in the integrand of the
partition function of superconformal theories on
squashed $\IS^5$ \cite{lv,Imamura:2012bm}. However, as emphasized in
\cite{lv}, the triple sine functions
cannot be simply identified with the 
partition functions of refined topological strings.
Instead, Lockhart and Vafa proposed that
the partition function of non-perturbative 
refined topological strings is defined
by a triple product of refined partition functions:
\begin{align}
 Z_{\rm np}^{({\rm LV})}=
Z_{\rm ref}({\bf T},\tau_1+1,\tau_2)
Z_{\rm ref}\left(\frac{{\bf T}}{\tau_1},\frac{1}{\tau_1},\frac{\tau_2}{\tau_1}+1\right)
Z_{\rm ref}\left(\frac{{\bf T}}{\tau_2},\frac{\tau_1}{\tau_2}+1,\frac{1}{\tau_2}\right),
\label{ZLVtriple}
\end{align}
where the refined partition function is given by the exponential of (\ref{refBPS}) 
\begin{align}
Z_{\rm ref}({\bf T},\tau_1,\tau_2)
=\prod_{{\bf d}}\prod_{j_L,j_R}
\prod_{m_L=-j_L}^{j_L}\prod_{m_R=-j_R}^{j_R}\prod_{n_1,n_2=0}^\infty
(1-q_1^{m_L+m_R+n_1+\frac{1}{2}}q_2^{-m_L+m_R-n_2-\frac{1}{2}}\re^{-{\bf d}\cdot {\bf T}})
^{N^{{\bf d}}_{j_L,j_R}},
\label{Zref}
\end{align}
and we have denoted, as in \cite{lv}, 
\be
\epsilon_{1,2}= 2 \pi \ri \tau_{1,2}.  
\ee
Notice that the first factor in (\ref{ZLVtriple}) is the perturbative, refined topological partition function, while the second and third factors involve non-analytic terms in the 
coupling constants $\tau_{1,2}$. This proposal is similar to (\ref{FnpFref}) and \eqref{ourFnp}: in both of them, the non-perturbative corrections involve a refined topological string on a different ``slice" of the $\tau_1-\tau_2$ space, the coupling constants are inverted, and the corrections are of the form $\re^{-{\bf T}/\tau_i}$. In order to make a more detailed comparison to our results, we should consider a particular case of (\ref{Zref}) in which the perturbative sector is the ordinary topological string with $\epsilon_1=-\epsilon_2$.

As discussed in \cite{lv}, 
some of the factors in \eqref{ZLVtriple}
can be moved to the denominator
by analytic continuation. Combining this with the symmetry
of the refined partition function 
\be
Z_{\rm ref} ({\bf T},\tau_1,\tau_2)=Z_{\rm ref} ({\bf T},-\tau_2,-\tau_1),
\ee
we can rewrite \eqref{ZLVtriple} as
\begin{align}
 Z_{\rm np}^{({\rm LV})}=
Z_{\rm ref}({\bf T},\tau_1+1,\tau_2)
\frac{Z_{\rm ref}\left(\frac{{\bf T}}{\tau_1},\frac{1}{\tau_1},\frac{\tau_2}{\tau_1}+1\right)}
{Z'_{\rm ref}\left(\frac{{\bf T}}{\tau_2},\frac{1}{\tau_2},-\frac{\tau_1}{\tau_2}-1\right)},
\label{ZLVprime}
\end{align}
where the prime signifies that $SU(2)_L$ and $SU(2)_R$ are
exchanged.

To reproduce our expression \eqref{ourFnp}
in the single coupling case,
we have to  modify \eqref{ZLVprime} by changing some of the signs
in the denominator 
\begin{align}
  Z_{\rm np}=
Z_{\rm ref}({\bf T},\tau_1+1,\tau_2)
\frac{Z_{\rm ref}\left(\frac{{\bf T}}{\tau_1},\frac{1}{\tau_1},\frac{\tau_2}{\tau_1}+1\right)}
{Z'_{\rm ref}\left(-\frac{{\bf T}}{\tau_2},-\frac{1}{\tau_2},-\frac{\tau_1}{\tau_2}-1\right)}.
\label{ZnpZref}
\end{align}
We should stress that
our proposal \eqref{ZnpZref} is not the unique
expression which reduces to \eqref{ourFnp}
in the single coupling case.
We choose \eqref{ZnpZref} just as a simple modification
of the proposal \eqref{ZLVprime} in \cite{lv}.

It is natural to identify
the first factor of \eqref{ZnpZref}
as the worldsheet instanton corrections and
the second factor as the ``membrane instanton'' corrections. Note that the extra sign for the worldsheet instanton
\begin{align}
 \re^{2\pi \ri(m_L+m_R+\frac{1}{2})}=(-1)^{2j_L+2j_R-1}
\end{align}
naturally appears from the shift $\tau_1\rightarrow\tau_1+1$
in the first factor of \eqref{ZnpZref}.
It is interesting that this sign was introduced in \cite{lv}
by a very different argument from ours.
In our case, this sign was introdued
from the requirement of pole cancellation, while 
the argument of \cite{lv}
is based on the consideration of spin structure.

Now let us specialize to the one parameter case
$\tau_1+\tau_2=0$. 
Then, the first factor of \eqref{ZnpZref} becomes the unrefined topological
string partition function with the discrete B-field turned on,
as we have seen in the previous subsection.
For the second factor of 
\eqref{ZnpZref}, due to the
shift of parameter by one, 
setting 
$\tau_1+\tau_2=0$ amounts to taking the NS limit.
Also, the
derivative of 
$F_{\rm NS}$ in \eqref{ourFnp}
can be reproduced by taking the limit
$\tau_1+\tau_2\rightarrow0$ carefully. Namely, we set
\begin{align}
\tau_1=\lambda_s+\varepsilon,\quad\tau_2=-\lambda_s,
\end{align}
and take the limit $\varepsilon\to 0$ at the end of computation.
Recall that in the NS limit only 
the diagonal $SU(2)_{\rm diag}\subset SU(2)_L\times SU(2)_R$
couples non-trivially to $q_1$ and hence
the exchange of $SU(2)_L$ and $SU(2)_R$ does not matter in the NS limit.
Thus, 
the
log of the second factor of \eqref{ZnpZref} 
 becomes
\begin{align}
&\lim_{\varepsilon\rightarrow0}\left[
F_{\rm ref}\left(\frac{{\bf T}}{\lambda_s+\varepsilon},
\frac{1}{\lambda_s+\varepsilon},\frac{\varepsilon}{\lambda_s+\varepsilon}\right)
-F'_{\rm ref}\left(\frac{{\bf T}}{\lambda_s},\frac{1}{\lambda_s},
\frac{\varepsilon}{\lambda_s}\right)
\right]
\nn
=&\lim_{\varepsilon\rightarrow0}
\left[\frac{\lambda_s+\varepsilon}{2\pi\ri\varepsilon}F_{\rm NS}\left(\frac{{\bf T}}{\lambda_s+\varepsilon},\frac{1}{\lambda_s+\varepsilon}\right)
-\frac{\lambda_s}{2\pi\ri\varepsilon}F_{\rm NS}\left(\frac{{\bf T}}{\lambda_s},\frac{1}{\lambda_s}\right)\right]\nn
=&\,\frac{1}{2\pi\ri}\frac{\partial}{\partial \lambda_s}\left[\lambda_sF_{\rm NS}\left(\frac{{\bf T}}{\lambda_s},\frac{1}{\lambda_s}\right)\right],
\end{align}
which shows that \eqref{ZnpZref} almost reduces to
our conjectured form of the non-perturbative free energy 
for the single coupling case \eqref{ourFnp}. One difference with our proposal in the previous subsection 
is that the K\"ahler parameters should be promoted to ``effective" K\"ahler parameters 
incorporating the effects of the bound states, as well as the non-perturbative B-field. In that sense, the proposal of \cite{lv} seems to lead to a complete 
factorization between perturbative and non-perturbative sectors and misses the contribution of bound states. It would be interesting to understand 
the modification \eqref{ZnpZref} in the context of \cite{lv}. It might be due to the fact that the ordinary topological string is a degenerate case of the formalism in \cite{lv}, 
since $\tau_{1,2}$ are aligned. On top of that, our expansion of the free energy 
corresponds to $\lambda_s$ real and positive, therefore ${\rm Im}\, \tau_1={\rm Im}\, \tau_2=0$, and one should be 
more careful with the convergence properties of the triple sine functions. 

Notice that we could regard \eqref{ZnpZref} as an appropriate 
generalization of our proposal to the case of refined topological strings, since it essentially reduces to our proposal in the case 
of standard, unrefined strings. However, since for general refined strings we do not have a notion of quantum A-period, 
it is not obvious how to extend our proposal for the effective K\"ahler parameters (\ref{teffgen}) to the general, refined case. 

\sectiono{Conclusions and prospects for future work}

In this paper we have determined the complete non-perturbative expansion of the partition function of ABJM theory on the three-sphere. The resulting picture is beautiful and appealing: worldsheet instanton corrections 
are determined by the standard topological string on local $\IP^1 \times \IP^1$. Membrane instanton corrections are determined 
by the refined topological string on the same CY, and in the Nekrasov--Shatashvili limit. 
Mathematically, this means that the large $\mu$ expansion of the grand potential (which is a large $N$ expansion), as determined by the TBA equations appearing in the Fermi gas approach, agrees with the large radius expansion of the quantum periods. Since this expansion can be computed at finite $k$, we also have an efficient method to calculate the large $N$ expansion at fixed $k$, as required in the M-theory expansion. 

Although we have overwhelming evidence for this equivalence, it remains a conjecture. It would be very interesting to prove it in order to establish our claim. Mathematically, this would provide an interesting link between 
the TBA formulation of the Fermi gas and the problem of calculating the quantum periods of this local CY. Given the deep relationship between refined topological strings and integrable systems \cite{ns}, this link is maybe not that surprising, but its clarification could lead to additional insights on this relationship. Notice that the TBA equations appearing in the Fermi gas approach are very close to those calculating indices in two-dimensional, ${\cal N}=2$ theories \cite{cfiv}. Since the NS limit of the refined topological string also leads naturally to a ${\cal N}=2$ theory in two dimensions \cite{ns}, the connection might be due to this common two-dimensional origin. 

The answer we have found for the non-perturbative membrane effects gives also the full set of non-perturbative corrections to the free energy of topological string theory on local $\IP^1 \times \IP^1$, along the diagonal direction. 
In doing this we assumed that the non-perturbative partition function of local $\IP^1 \times \IP^1$ is given by the ABJM matrix model. 
This is certainly natural from the point of view of the duality between this topological string and Chern--Simons theory on $\IR \IP^3$ \cite{akmv-cs}, and 
the relation between the Chern--Simons matrix model on $\IR \IP^3$ and the ABJM matrix model \cite{mp}. Based on this result, we have also made a proposal for the non-perturbative effects of topological string theory on arbitrary, local CY manifolds. It would be certainly important to test if this proposal is true. One possible strategy is to consider the class of general 
$A_{N-1}$ fibrations over $\IP^1$, which have Chern--Simons/matrix model descriptions \cite{akmv-cs}, and try to compute the non-perturbative effects in these models in a similar way. 

We have also pointed out that our result bears some resemblance to the proposal of \cite{lv}, although some important aspects of our concrete, 
first-principles calculation (like the presence of bound states) do not seem to be captured by 
the proposal in \cite{lv}. The moral lesson of \cite{lv} seems to be that the triple sine function, or some modification thereof, 
has the right properties to encode the perturbative topological string free energy as well as its 
non-pertubative corrections. It would be interesting to understand better the relationship between our proposal and the approach in \cite{lv}. 

Finally, it would be interesting to study non-perturbative effects in more general Chern--Simons--matter theories. 
One obvious, simple generalization of this work is ABJ theory \cite{abj}, but one could consider the more general class 
of $\CN=3$ theories which can be formulated as free Fermi gases \cite{mp-fermi}.

\section*{Acknowledgements}
We would like to thank Flavio Calvo, Guglielmo Lockhart, Atsushi Narukawa, Boris Pioline, 
Soo-Jong Rey, Cumrun Vafa and Stefan Vandoren for very useful 
conversations and correspondence. We are particularly thankful to Albrecht Klemm and Daniel Krefl for the generous explanations of their work. 
S.M. would like to thank the Yukawa Institute for Theoretical Physics at Kyoto University for hospitality. 
M.M. and K.O. would like to thank the Kobayashi--Maskawa Institute at Nagoya University for hospitality during the initial 
stage of this collaboration. K.O. would also like to thank Seoul National University for hospitality.
The work of M.M. is supported in part by the Fonds National Suisse, subsidies 200020-126817 and 
200020-137523. The work of K.O. is supported in part by JSPS Grant-in-Aid for Young Scientists
(B) 23740178. 

\appendix
\sectiono{Quantum $A$-periods from the TBA system}
In this appendix, we improve the analyis of the TBA system done in \cite{cm} in order to extract the coefficients $a_\ell(k)$ in closed form. 
As shown in \cite{cm}, the TBA equations (\ref{tba-k}) can be written as 
\be
\label{semitba}
\ba
1+ \eta^2(x) &= R_+ \left( x +{\pi \ri k \over 2}\right) R_+ \left( x -{\pi \ri k \over 2}\right) \exp\left\{ U \left( x +{\pi \ri k \over 2}\right) +U \left( x -{\pi \ri k \over 2}\right) \right\},\\
-z R_+(x)&=  \eta \left( x +{\pi \ri k \over 2}\right) +\eta \left( x -{\pi \ri k \over 2}\right). 
\ea
\ee
where $U(x)$ is given by (\ref{ux}). We can now plug the second equation into the first one and obtain a single equation for $\eta$, as in \cite{tw}. If we introduce the variables
\be
X=\re^x, \qquad q=\re^{ \pi \ri  k }, \qquad \lambda={1\over z^2},
\ee
as well as 
\be
\widetilde \eta = \ri \eta,
\ee
this equation reads, 
\be
1- \widetilde \eta^2(x) + \lambda \left[ \widetilde \eta  \left(q X \right) +\widetilde \eta\left(X\right) \right]\left[ \widetilde\eta  \left( q^{-1} X \right) +\widetilde \eta\left(X\right) \right]\left( X+ X^{-1} + q^{1/2}+ q^{-1/2} \right)=0.
\ee
This can be solved in a power series in $\lambda$, 
\be
\widetilde \eta (X)= \sum_{n\ge 0} \eta_n(X) \lambda^n=1+2 \left( X+ X^{-1} + q^{1/2}+ q^{-1/2} \right) \lambda+ \cdots
\ee
From the equality \cite{cm}
\be
\label{int-R}
{1\over 4 \pi k} \int_{-\infty}^{\infty} R_+(x) \, \rd x = \sum_{\ell\ge 1}  \left[  \left( 2 \pi \ri \log z - \pi^2 \right)  \ell a_\ell(k) + \pi \ri \left(\ell b_\ell(k) - a_\ell(k) \right)  \right] z^{-2\ell-1} 
\ee
it follows that the coefficient $a_\ell (k)$ can be obtained from the real part of the integral of $R_+(x)$, which can be computed by contour deformation as a residue at infinity. Since
\be
R_+(X)= -\ri \lambda^{1/2}\omega(X),
\ee
where
\be
\omega(X)=  \widetilde \eta (q^{1/2} X) + \widetilde \eta (q^{-1/2} X)=\sum_{n \ge 0} \omega_n(X) \lambda^n, 
\ee
we conclude that
\be
k a_\ell (k)= -{1\over 2 \pi^2 \ell} {\rm Res}_{X=0}\,  \omega_\ell(X). 
\ee
This gives a very efficient way to compute the coefficients $a_\ell(k)$ appearing in the grand potential $J(\mu, k)$, 
which can then be compared to the quantum $A$-periods of local $\IP^1 \times \IP^1$. 
 
 \sectiono{Quantum mirror map}
As we have shown in section 3.3, 
the relation between $\mu$ and
$\mu_{\rm eff}$ can be interpreted as a quantum mirror map. 
As discussed in \cite{hmo2},
when one inverts the the relation between $\mu$ and
$\mu_{\rm eff}$ as
\begin{align}
\label{e}
\mu=\mu_{\rm eff}
+\frac{1}{C(k)}\sum_{\ell=1}^\infty e_\ell(k)\re^{-2\ell\mu_{\rm eff}},
\end{align}
one finds that the coefficients $e_\ell(k)$ take a simpler form than the
original $a_\ell(k)$.
We will see that this map can be
expressed in terms of some integer invariants.

Using the quantum A-period in
\eqref{q-Aper},
we find that the quantum mirror map of
local $\mathbb{P}^1\times \mathbb{P}^1$ \eqref{qmm}
has a multi-covering
structure
\begin{align}
 \frac{1}{2}\log\frac{Q_I}{z_I}
=\sum_{n=1}^\infty\sum_{d_1,d_2}\sum_j (-1)^{(n-1)d}\mathcal{N}_j^{d_1,d_2}
\chi_j(q^n)\frac{(Q_1^{d_1}Q_2^{d_2})^n}{n},
\label{mirror-cover}
\end{align}
where $d=d_1+d_2$ denotes the total degree and $\mathcal{N}_j^{d_1,d_2}$ are integer numbers.
This implies
that the factor $(Q_I/z_I)^{\frac{1}{2}}$ appearing in the open flat coordinate,
which represents the instanton corrections to the disk amplitude \cite{AKV},
has the following product expression
\begin{align}
 \left(\frac{Q_I}{z_I}\right)^{\frac{1}{2}}
=\prod_{d_1,d_2}\prod_j\prod_{m=-j}^j\Big(1+(-1)^{d-1}q^{2m}Q_1^{d_1}Q_2^{d_2}\Big)^{
(-1)^{d-1}\mathcal{N}_j^{d_1,d_2}}.
\end{align}
The integers  $\mathcal{N}_j^{d_1,d_2}$
might be interpreted as a refined version of the number of BPS
states 
in the presence of a
D-brane domain wall \cite{AKV}, in the NS limit.
These invariants are symmetric
in $d_1,d_2$:
$\mathcal{N}_j^{d_1,d_2}=\mathcal{N}_j^{d_2,d_1}$.
The first few non-zero values are given by
\begin{align}
 \mathcal{N}_j^{1,n}=\delta_{j,\frac{n}{2}}~~(n\geq0),\qquad
\mathcal{N}_1^{2,2}=1,\quad
\mathcal{N}_{\frac{3}{2}}^{2,2}=4,\quad
\mathcal{N}_2^{2,2}=1.
\end{align}
We also observed that
$M^{d_1,d_2}(q)=\sum_j\mathcal{N}_j^{d_1,d_2}\chi_j(q)$ can be factorized
by $\chi_{\frac{d-1}{2}}(q)$ and the remaining part
has an interesting pattern of coefficients:
\begin{align}
 &M^{2,2}=\chi_{\frac{3}{2}}(\chi_{\frac{1}{2}}+4\chi_0),\quad\quad
M^{2,3}=\chi_{2}(\chi_1+4\chi_{\frac{1}{2}}+8\chi_0),\nn
&M^{2,4}=\chi_{\frac{5}{2}}(\chi_{\frac{3}{2}}+4\chi_1+8\chi_{\frac{1}{2}}+12\chi_0),\quad
M^{3,3}=\chi_{\frac{5}{2}}(\chi_2+4\chi_{\frac{3}{2}}+12\chi_1+24\chi_{\frac{1}{2}}
+30\chi_0).
\end{align}
Here we have suppressed the argument $q$.

By specializing to the ABJM case
$Q_{1,2}=q^{\pm\frac{1}{2}}\re^{-2\mu_{\rm eff}}$,
we find that 
$e_\ell(k)$ in \eqref{e}
is written in terms of the integer invariants 
$\mathcal{N}_j^{d_1,d_2}$ as
\begin{align}
\frac{1}{C(k)}e_\ell(k)=\sum_{d|\ell}\sum_{d_1+d_2=d}
\sum_j\frac{d}{\ell}(-1)^{\ell-d}
\mathcal{N}_j^{d_1,d_2}\chi_j(q^\frac{\ell}{d})
q^{\frac{\ell(d_1-d_2)}{2d}}.
\end{align}

\end{document}